\documentclass[aps,prb,twocolumn]{revtex4} %,preprint

\usepackage{graphicx}% Include figure files
\usepackage{dcolumn}% Align table columns on decimal point
\usepackage{bm}% bold math

\newcommand{\comment}[1]{}
\begin{document}
%\preprint{}   % Preprint number in upper right corner

\title{How to Measure Forces
when the Atomic Force Microscope
shows Non-Linear Compliance}

%\author{}
%\thanks{}

\author{Phil Attard}
%\affiliation{\protect\texttt{phil.attard1@gmail.com}}

\date{16 November, 2012. phil.attard1@gmail.com}

\begin{abstract}
A spreadsheet algorithm is given
for the atomic force microscope
that accounts for non-linear behavior
in the deflection of the cantilever
and in the photo-diode response.
In addition,
the data analysis algorithm takes into account
cantilever tilt,
friction in contact,
and base-line artifacts such as
drift,
virtual deflection,
and non-zero force.
These are important for accurate force measurement
and also for calibration of the cantilever spring constant.
%These corrections are also given for the linear case.
The zero of separation is determined automatically,
avoiding human intervention or bias.
The method is illustrated by analyzing measured data
for the  silica-silica drainage force and slip length.
\end{abstract}

%\keywords{}

\maketitle

%%%%%%%%%%%%%%%%%%%%%%%%%%%%%%%%%%%%%%%%%%%%%%%%%%%%%%%%%%%%%%%%%%%%%%%%%%%%%%%
%                                                                             %
                \section{Introduction}
%                                                                             %
%%%%%%%%%%%%%%%%%%%%%%%%%%%%%%%%%%%%%%%%%%%%%%%%%%%%%%%%%%%%%%%%%%%%%%%%%%%%%%%

Scanning probe microscopy
has revolutionised surface science
by enabling the production of images of surfaces
with molecular resolution.
In one technique a topographic map is produced
based on the height adjustment of a piezo-drive
required to maintain a constant deflection of the cantilever probe
during a raster scan.
The original atomic force microscope
used tunneling currents to detect the deflection
of the cantilever.\cite{Binnig86}
This was soon modified to use a light lever
to detect the deflection,
which had the advantage of not requiring vacuum conditions.\cite{Marti88,Meyer88}
It also allowed quite large deflections to be measured,
and different imaging techniques to be developed,
such as constant height imaging.
For such techniques to be quantitative,
the light lever signal (in volts) has to be converted
into the deflection of the cantilever (in nanometers).

This issue of quantitatively calibrating the light lever
received added impetus with the further modification
of the atomic force microscope
in what has come to be called colloid probe force microscopy.\cite{Ducker91}
In this technique a colloid sphere of measured radius
($R \approx$ 10--20$\,\mu$m)
is glued to the end of the cantilever spring
instead of the sharp tip used for imaging.
The object is to measure the so-called surface force between
the substrate and the probe as a function of separation.
The surface force is
just the spring constant times the cantilever deflection,
and so if the light lever is properly calibrated,
the measurement may be performed with molecular resolution.

The light lever itself is usually made from
an optical beam reflecting off the back of the cantilever spring
onto a split photo-diode.
A change in angle of the cantilever,
due, for example, to a change in the force on the probe,
causes the light beam to move across the face of the photo-diode.
The consequent change in voltage difference between the two halves
is measured  and taken to be proportional to the change in cantilever angle.
By using the piezo-drive to press the cantilever against
the hard substrate,
the proportionality constant is obtained
as the slope of the voltage versus distance signal.

%%%%%%%%%%%%%%%%%%%%%%%%%%%%%%%%%%%%%%%%%%%%%%%%%%%%%%%%%%%%%%%%%%
\begin{figure}[t!]
\centerline{
\resizebox{8.5cm}{!}{ \includegraphics*{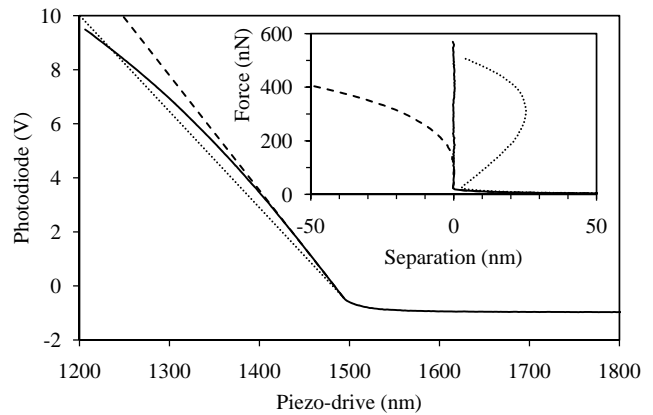} } }
% From My Documents\Papers\Current\nonlinear AFM\nl-SiSi2.xlsx:Fig1
\caption{\label{Fig:VvsZp}
The raw photo-diode voltage versus the piezo-drive displacement.
The solid curve is measured extension data,
the dashed line is the tangent to the contact region
at first contact, % $z_\mathrm{p} = 1493\,$nm,
and the dotted line uses the average slope in contact.
The inset shows the analysed force versus separation,
with the solid curve resulting from the non-linear analysis (see text)
and the dashed and dotted curves resulting from the conventional linear analysis
using the first contact slope and the average slope, respectively.
The source of the measured data is Ref.~\onlinecite{Zhu12};
a summary of the experimental details is given in Ref.~\onlinecite{expt}.
}
\end{figure}
%%%%%%%%%%%%%%%%%%%%%%%%%%%%%%%%%%%%%%%%%%%%%%%%%%%%%%%%%%%%%%%%%%

To make this clear mathematically,
let the  measured constant compliance slope be
\begin{equation}
\beta \equiv
\left.
\frac{\Delta V }{\Delta z_\mathrm{p}}
\right|_\mathrm{contact} .
\end{equation}
Here $V$ is the photo-diode voltage
and $z_\mathrm{p}$ is the piezo-drive position.
%As is explained in detail below,
%it is necessary to distinguish between piezo-drive extension (approach)
%and retraction (withdrawal).
Letting $V_\mathrm{b}$ be the base-line voltage far from contact,
the surface force is
\begin{equation}
F(z_\mathrm{p})
=
- k_0 \beta ^{-1}
[ V(z_\mathrm{p}) - V_\mathrm{b} ] ,
\end{equation}
where $k_0$ is the cantilever spring constant.
The force goes to zero at large separations,
which is the base-line region.
This result invokes the fact that  in hard contact
the change in tip position
is equal and opposite to the change in piezo-drive position, \cite{NB1}
$ \Delta z_\mathrm{t} = -\Delta z_\mathrm{p}$,
and the force \emph{on} the cantilever is $ F = k_0 z_\mathrm{t}$.
(The simple presentation given here ignores a number of important linear effects
such as base-line drift, friction in contact, cantilever tilt,
and non-negligible base-line force.
These and non-linear effects
will be included in the more sophisticated
linear and non-linear analysis below.)
The separation is
\begin{eqnarray}
h & = &
z_\mathrm{p} +  z_\mathrm{t} + \mbox{const}
\nonumber \\ & = &
z_\mathrm{p} +  k_0 ^{-1}F^\mathrm{ext}(z_\mathrm{p})
+ \mbox{const} ,
\end{eqnarray}
where the constant is chosen so that $h=0$ in contact.

It ought to be clear from the above how important the calibration factor
$\beta$,
which is the constant compliance slope,
is to the quantitative measurement of surface forces
with the atomic force microscope.
However it is not unusual for the photo-diode voltage
versus piezo-drive position curve to be non-linear
in the contact region.
A typical example is shown in Fig.~\ref{Fig:VvsZp}.
It is emphasised that the data in the figure were obtained
for hard surfaces and so the curvature evident
is not due to elastic deformation of the probe or substrate.
%All of the data and analysis in this paper are for rigid surfaces.
A curved constant compliance region such as that in Fig.~\ref{Fig:VvsZp}
creates several problems.
At a minimum, the calibration factor $\beta$ is not unique;
the slope depends upon where in the contact region it is measured,
which introduces quantitative uncertainty into the linear analysis,
(compare, for example,
the result that uses the slope at first contact (dashed line and curve)
with the result that uses the average slope (dotted line and curve)).
Ambiguity also arises because the contact behavior
differs between extension and retraction (not shown).
Worse, a non-linear contact region contradicts
the fundamental assumption of a linear response
and raises questions about the conventional linear analysis
that is used to quantify surface force measurements.

The inset to Fig.~\ref{Fig:VvsZp} graphically illustrates
the problems with the linear analysis.
It can be seen that it gives grossly unphysical non-zero separations
in the contact region.
In this case the surfaces are known to be rigid so the results
is unambiguously unphysical.
In other cases, where the surfaces are either of unknown rigidity
or known to be soft, this artefact of the linear analysis
would be misinterpreted as the elastic deformation of the material.
Not only the elasticity
but any useful physical information
in the contact region is precluded by the linear analysis.
As will be demonstrated below, the non-linear analysis
can be used to  obtain reliable values for properties like
the friction coefficient, roughness, and topography of the
contact region.

It is not only in contact that the linear analysis can fail.
If a large surface force is present,
then the linear analysis of the data introduces quantitative errors
into the values of the surface force in the non-contact region.
This can be a particular problem if one requires reliable and accurate
surface force measurements,
or if one seeks small changes in forces with control parameters,
or if one needs to quantify second order effects.
In all these cases the linear analysis can be unsuitable,
depending upon the extent of the non-linearity and the magnitude of the forces.

The are two possible physical origins
of the non-linearity displayed in Fig.~\ref{Fig:VvsZp}.
The first possibility is that the cantilever deflection becomes non-linear
over the relatively large range of the contact region.
By cantilever non-linearity is meant that the four relevant quantities
(tip position, deflection, angle deflection, and force)
are not linearly proportional to each other.
The second possibility is a non-linear relationship
between the change in photo-diode voltage and the change in cantilever angle.
Since the change in voltage difference in the split photo-diode
depends only on that part of the optical beam currently crossing
the boundary
(assuming uniform sensitivity of the photo-diode;
any spatial variability in the sensitivity will contribute further
to non-linearities),
any variability in the spatial intensity or width of the beam
(e.g.\ circular or elliptical cross-section, Gaussean intensity distribution)
will give non-linear effects.

%%%%%%%%%%%%%%%%%%%%%%%%%%%%
\subsubsection{Contents}

For the case of the rectangular cantilever,
a complete linear analysis is given in \S \ref{Sec:Lin-Analysis}.
In \S\ref{Sec:Lin-tilt},
the effects of base-line drift, virtual deflection,
cantilever tilt, friction in contact, and non-negligible base-line force
are accounted for.
These are often neglected in the conventional linear analysis.
A new result is an algorithm for determining the zero of separation,
\S\ref{Sec:z0-lin}.
The effective spring constant that must be used
in the linear analysis and its relation
to the cantilever spring constant
is given in \S\ref{Sec:keff}.
In \S\ref{Sec:nl-cant} are given
the non-linear equations for a rectangular cantilever
that  determine the deflection,
deflection angle, vertical position, and applied force,
taking into account tilt and friction.
In \S\ref{Sec:nl-anal}  an algorithm suitable for spreadsheet use
is given for the analysis of experimental data
(i.e.\ the conversion from raw voltage versus piezo-drive position
to force versus separation)
in the case of non-linear behavior.
The non-linearities can arise either from the non-linear cantilever
deflection or the non-linear photo-diode response, or both.
For the non-linear photo-diode case,
the non-linearity is characterized by the measurement itself
and  it is not necessary to know the details of the source of the non-linearity.
This is fortunate because unlike rectangular cantilevers,
these vary between different models of the atomic force microscope.
In \S\ref{Sec:nlcant-linpd}
the case of a linear photo-diode and non-linear cantilever
is explored numerically for the case shown in Fig.~\ref{Fig:VvsZp},
and it is concluded that the non-linear cantilever deflection
is insufficient to account for the measured non-linear effects.
In \S\ref{Sec:LinCant-nlpd},
summarized are
the equations for a linear cantilever and a non-linear photo-diode,
which are somewhat simpler than the dual non-linear case.
In \S\ref{Sec:results},
these are applied to measured atomic force microscope data
for the drainage force at several drive velocities.
Results for %the friction coefficient,
the slip length
and the drainage adhesion are obtained.
The quantitative and qualitative differences
between the linear and the non-linear analysis
of the experimental data are shown.

%In \S III the non-linear response of the photo-diode is examined.
%In this case due to several unknowns
%(the shape of the beam, the beam path, the effect of mirrors and lenses,
%the electronic performance of the photo-diode)
%it is not possible to give an analytic formula for the non-linearities.
%Also, these vary between different atomic force microscope models;
%unlike the universal results for rectangular cantilevers,
%any analytic results for photo-diode non-linearity
%for one particular model would be of no use for another model.

%%%%%%%%%%%%%%%%%%%%%%%%%%%%%%%%%%%%%%%%%%%%%%%%%%%%%%%%%%%%%%%%%%%%
%
\section{Linear Analysis}
%
%%%%%%%%%%%%%%%%%%%%%%%%%%%%%%%%%%%%%%%%%%%%%%%%%%%%%%%%%%%%%%%%%%%%
\label{Sec:Lin-Analysis}

%%%%%%%%%%%%%%%%%%%%%%%%%%%%%%%%%%%%%%%%%%%%%%%%%%%%%%%%%%%%%%
\subsection{Horizontal Cantilever, No Friction}
\label{Sec:lin-horiz-mu=0}

%%%%%%%%%%%%%%%%%%%%%%%%%%
\subsubsection{Deflection}

The bending of a cantilever beam under the influence
of fores and torques
is one of the classic problems of the theory of elasticity.
A beam of length $L_0$,
and with a deflection $x$
and angular deflection $\theta$
has stored elastic  energy\cite{Southwell36}
\begin{equation}
U(x,\theta) =
\frac{2B}{L_0^3} \left[ 3 x^2 - 3 x L_0 \theta + L_0^2 \theta^2 \right] .
\end{equation}
The elastic parameter  $B \equiv  EI$ depends upon
Young's modulus and the geometric second moment of the
beam; it will be related to the spring constant of the cantilever beam below.
Differentiating with respect to the deflection gives the
force exerted \emph{on} the end of the beam,
\begin{equation}
F \equiv \frac{\partial U(x,\theta)}{\partial x}
=
\frac{2B}{L_0^3} \left[ 6 x - 3 L_0 \theta  \right] .
\end{equation}
and differentiating with respect to angle gives the torque
\begin{equation}
\tau \equiv \frac{\partial U(x,\theta)}{\partial \theta}
=
\frac{2B}{L_0^3} \left[- 3 L_0 x  + 2 L_0^2 \theta \right] .
\end{equation}
These assume that the beam is in equilibrium with the applied
forces and torques.

Inverting these equations gives the standard expressions
for the deflection and the angle in terms of the applied
force and torque,\cite{Southwell36}
\begin{equation} \label{Eq:x(F,tau)}
x
=
\frac{1}{2B} \left[\frac{2}{3} L_0^3 F + L_0^2  \tau \right] ,
\end{equation}
and
\begin{equation}  \label{Eq:q(F,tau)}
\theta
=
\frac{1}{2B} \left[ L_0^2 F+ 2 L_0 \tau  \right] .
\end{equation}

%%%%%%%%%%%%%%%%%%%%%%%%%%%%%%%%
\subsubsection{Spring Constant}

One has to be a little cautious about assigning a spring constant.
The above equations refer to a free, horizontal cantilever beam,
and now the spring constant for such a beam with normal force and zero torque
(free deflection) will be given.
It is emphasized that this cannot be applied to the atomic force microscope
without modification because in that case
the cantilever beam is tilted,
the force is not entirely normal to the beam,
and there are non-zero torques due to this and due to friction.
This case will be handled shortly.

For the horizontal beam with zero torque, $\tau = 0$,
the deflection is linearly proportional to the applied force,
$x = [ L_0^3 /3B ] F$,
from which one can identify the cantilever spring constant as
\begin{equation}
k_0 \equiv 3 B/L_0^3 .
\end{equation}
In this paper the elastic parameter $B$ will be used,
as it is an intrinsic property of the cantilever.
Usually (but not always; see \S\ref{Sec:keff} below)
any quoted or measured spring constant
is the horizontal, free spring constant in the above sense,
and so this equation can be used to convert from $k_0$ to $B$.

For free deflection, $\tau = 0$,
the angular deflection is linearly proportional to the deflection,
\begin{equation}
\theta = \frac{3}{2 L_0} x .
\end{equation}
This particular proportionality constant
only holds for the free, horizontal cantilever.
For the free tilted cantilever,
the two remain linearly proportional to each other,
but a different constant applies,
as is derived below.

The linear proportionality of deflection angle $\theta$
and deflection $x$ underlies the linear analysis
of atomic force microscope.
The light lever is assumed to give a change in voltage
that is linearly proportional to the deflection angle,
$\gamma \equiv \Delta V/\Delta \theta$.
Since the deflection angle, deflection, and force,
are all linearly proportional to each other by the above two equations,
then one can get the force from the measured change in voltage,
\begin{equation}
F = k_0 x = \frac{2 k_0 L_0}{3} \theta = \frac{2 k_0 L_0}{3 \gamma}  \Delta V.
\end{equation}
If the measured gradient of the photo-diode signal in contact is
$\beta \equiv \Delta V/\Delta z_\mathrm{p}
= -\Delta V/\Delta x$,
then
\begin{equation}
\gamma = \frac{- 2 L_0}{3} \beta .
\end{equation}
Hence $F = -k_0  \beta^{-1}  \Delta V$.
This is the conventional linear analysis
for extracting the force from of atomic force microscope measurements.
It ought be clear it assumes a horizontal cantilever with no torque,
neither of which assumption holds in practice.

%%%%%%%%%%%%%%%%%%%%%%%%%%%%%%%%%%%%%%%%%%%
\subsubsection{Spring Constant Calibration}

One of the most important issues in measuring forces
with the atomic force microscope is the determination of the spring constant.
This is usually the largest source of systematic error.
The common thermal calibration procedure\cite{Higgins06}
that is often built into the software of the atomic force microscope
gives erroneous results,
with a systematic overestimate of the spring constant
of 15\%--30\%.\cite{Higgins06,Attard06a}
The source of the error in the derivation has been identified
and a more reliable thermal calibration method has been given.\cite{Attard06a}
(The correct thermal calibration formula for the cantilever spring constant
is given in Eq.~(\ref{Eq:k-therm}) below.)
An even more accurate and reliable way of determining
the spring constant
is to use the long range hydrodynamic drainage force.
\cite{Zhu11a,Zhu11b,Craig01}
In general the drainage force is known exactly,
at least in the large separation regime
\begin{equation}
F_\mathrm{drain}(h) = \frac{-6 \pi \eta R^2 \dot z_\mathrm{p}}{h} ,
\end{equation}
where $\eta$ is the viscosity.
It is permissible to use the piezo-drive velocity $\dot z_\mathrm{p}$
rather than the rate of change of change of separation in this
because the deflection is small and its rate of change is negligible
at large separations.
The correct spring constant gives agreement between
this and the measured force at long range.

It should be noted that if this method of calibration is used
in conjunction with the conventional analysis of the measured data
(linear calibration, horizontal cantilever),
then the spring constant that results is the effective spring constant
$k_\mathrm{eff}$ rather than the intrinsic cantilever spring constant $k_0$.
These are defined in the full analysis for the tilted cantilever with friction
that is treated next.

% \newpage $\,$ \newpage

%%%%%%%%%%%%%%%%%%%%%%%%%%%%%%%%%%%%%%%%%%%
\subsection{Tilted Cantilever with Friction}  \label{Sec:Lin-tilt}

\subsubsection{Model}

%%%%%%%%%%%%%%%%%%%%%%%%%%%%%%%%%%%%%%%%%%%%%%%%%%%%%%%%%%%%%%%%%%
\begin{figure}[t!]
\centerline{
\resizebox{8.5cm}{!}{ \includegraphics*{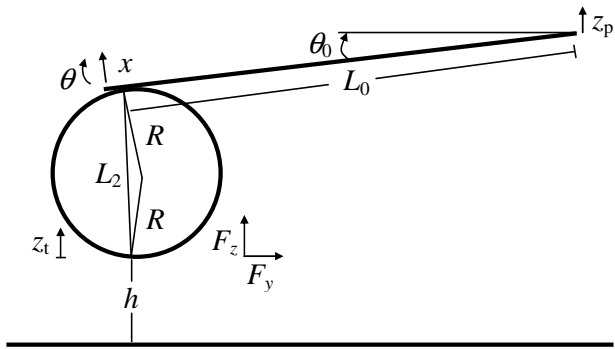} } }
% From My Documents\Papers\Current\nonlinear AFM\AFMGeom.doc
\caption{\label{Fig:AFMgeom}
Cantilever geometry in the atomic force microscope
(not to scale).
}
\end{figure}
%%%%%%%%%%%%%%%%%%%%%%%%%%%%%%%%%%%%%%%%%%%%%%%%%%%%%%%%%%%%%%%%%%

Following earlier work,\cite{Attard98,Attard99}
the cantilever and probe in the atomic force microscope
is modeled as in Fig.~\ref{Fig:AFMgeom}.
The key features are the fixed tilt angle of the cantilever, $\theta_0 < 0$,
such that the total angle of the cantilever is the sum of this
and the deflection angle, $\theta_\mathrm{tot} = \theta_0 + \theta$,
and the rigid lever arm, $L_2(\theta_\mathrm{tot})$,
which connects the cantilever to the point of application
of the normal surface force $F_z$
and the lateral friction force $F_y$, if present.
The force $F$ and torque $\tau$ on the cantilever
treated in the preceding section
are a function of these two forces, the lever arm, and the tilt angle,
as will now be derived.

Note that in the figure the piezo-drive is connected to the
base of the cantilever,
so that extension corresponds to $\dot z_\mathrm{p} < 0$
and retraction corresponds to $\dot z_\mathrm{p} > 0$.\cite{NB1}
The separation between the surfaces is
\begin{equation}
h \equiv z_\mathrm{p} + z_\mathrm{t} + z_0^\mathrm{ext},
\end{equation}
with the constant $z_0^\mathrm{ext}$ calculated to give zero separation
at first contact, as defined below.
In contact, $\Delta h =0$,
so that $\Delta z_\mathrm{t} = -\Delta z_\mathrm{p}$.\cite{NB1}

Trigonometric functions of the tilt angle occur frequently below
and so it is convenient to define fixed constants
\begin{equation}
C_0 \equiv \cos \theta_0
\mbox{, and }
S_0 \equiv \sin \theta_0 .
\end{equation}
Note that in the present geometry
for the atomic force microscope,
$S_0 \approx \theta_0 \approx -0.2$,
which is small in magnitude and negative in sign.
(All angles  here and throughout are measured in radians;
in degrees, $\theta_0 \approx -11^\circ $.)

%%%%%%%%%%%%%%%%%%%%%%%%%%%%%%%%
\subsubsection{Friction Force} \label{Sec:Friction}

In this work the lateral force will be taken to be due to friction
(in contact) and it will be taken to be linearly proportional to the load,
\begin{equation}
F_y = \left\{
\begin{array}{ll}
\mu F_z, & \mbox{extension, contact}, \\
-\mu F_z, & \mbox{retraction, contact}, \\
0 , & \mbox{non-contact}.
\end{array} \right.
\end{equation}
This is known as  Amontons' law.
As drawn in Fig.~\ref{Fig:AFMgeom},
on extension $\dot z_\mathrm{p} < 0$,
and so in contact on extension  $\dot z_\mathrm{t} > 0$.
This means that due to the tilt angle,
$\dot y < 0$ and $F_y > 0$.
Since in contact $F_z > 0$ (at least sufficiently far into contact),
this accounts for the sign of the first equality.
(In general the friction coefficient is positive.)
The opposite occurs on retraction
($\dot y > 0$ and $F_y < 0$), the second equality.
Out of contact there is no friction.

The assumption that the friction force
is linearly proportional to the load is a significant one.
This is certainly the classical model of friction,
at least at the macroscopic level.
There is evidence in the atomic force microscope literature
for\cite{Stiernstedt05}
and against\cite{Feiler00}
such an assumption.
The former data is perhaps the most convincing
as four independent measurements were made
(two different colloid probes, two different friction measurement methods).

Immediately after the initial contact on extension,
and at the beginning of the retraction branch in contact
(the turn point),
the probe is not moving at uniform velocity
and so the assumed form for the friction force will produce artifacts
at these points in the analyzed data.
Also, when the force at either first or last contact is non-zero,
the model gives a discontinuity in the friction force
and consequently a discontinuity in the surface force
that is also an artefact of the simple model.

All the following results will be given explicitly for extension in contact.
The retraction results may be obtained by the replacement
$\mu \Rightarrow -\mu$,
and the non-contact results may be obtained by the replacement
$\mu \Rightarrow 0$.
The superscripts `ext' (contact, $\mu > 0$),
`ret' (contact, $\mu < 0$), and `nc' (non-contact, $\mu = 0$)
will often be used to denote each of the three cases.
For the voltage, the piezo-drive position,
and the surface force, which is possibly velocity dependent,
the superscripts `ext' and `ret' will be used also in the non-contact situation.
The subscript `c' denotes a quantity in contact,
and the subscript `b' denotes a quantity in the base-line region
far from contact.

%%%%%%%%%%%%%%%%%%%%%%%%%%%%%%%%%%%%%%%%%%
\subsubsection{Linear Cantilever Equations} \label{Sec:lin-cant}

In Eqs (\ref{Eq:xFz}) and (\ref{Eq:QFz}) below,
non-linear expressions are derived
for the deflection, the angular deflection, and the force.
In the linear regime one can simply make the replacement
$\theta_\mathrm{tot} \Rightarrow \theta_0  $ to obtain
\begin{equation} \label{Eq:xFz-lin}
x =
\left\{
\frac{L_0^3}{3B}
\left[ C_0  + \mu S_0 \right]
+
\frac{ L_0^2  L_2 }{2B}
\left[ S_0
- \mu C_0 \right] \right\} F_z ,
\end{equation}
and
\begin{eqnarray} \label{Eq:QFz-lin}
\theta & = &
\left\{
\frac{L_0^2}{2B}
\left[ C_0   + \mu S_0 \right]
+
\frac{ L_0  L_2 }{B}
\left[ S_0
- \mu C_0  \right] \right\} F_z
\nonumber \\ & \equiv &
E^\mathrm{ext} F_z .
\end{eqnarray}
Here the length of the lever arm in the  undeflected state is
$L_2 \equiv L_2(\theta_0 ) = R \sqrt{ 2 +  2\cos \theta_0  } $.
In the linear regime,
the force, deflection, and deflection angle
are linearly proportional to each other.
In particular it is convenient to define the proportionality constant
between deflection and deflection angle from
\begin{eqnarray} \label{Eq:x(Q)}
x & = &
\frac{2 L_0^3 [ C_0 + \mu S_0 ] + 3 L_0^2 L_2 [ S_0 - \mu C_0 ]
}{ 3 L_0^2 [ C_0 + \mu S_0 ] + 6 L_0 L_2 [ S_0 - \mu C_0 ] }
\theta
\nonumber \\ & \equiv &
D^\mathrm{ext} \theta .
\end{eqnarray}
In general $L_2 \ll L_0$,
and this and the above results could be expanded to linear order
in $L_2/L_0$.
There is no great advantage in doing this.
%Recall that all of the above results are for extension in contact.
%The retraction results may be obtained by the replacement
%$\mu \Rightarrow -\mu$,
%and the non-contact results may be obtained by the replacement
%$\mu \Rightarrow 0$.

The vertical position of the tip depends upon the deflection,
the deflection angle, and the length of the lever arm,
Eq.~(\ref{Eq:ztQ}) below.
The linearized form of Eq.~(\ref{Eq:L2Q})
for the length of the lever arm is
$L_2(\theta_\mathrm{tot} ) =
L_2 - {R^2 S_0}\theta/{L_2} $.
For the case of a tipped cantilever,
$L_2(\theta_\mathrm{tot} )$ is equal to the length of the tip,
and there is no dependence on the deflection angle
(i.e.\ the term in $R^2$ may be set to zero).
Linearising  Eq.~(\ref{Eq:ztQ}) below for the vertical tip position  yields
\begin{eqnarray} \label{Eq:ztQ-lin}
z_\mathrm{t} & = &
C_0 x
+ \left[ L_2 S_0
- \frac{R^2 S_0^2 \theta_0}{L_2}
+ L_2  C_0 \theta_0 \right] \theta
\nonumber \\ & = &
\left\{ D^\mathrm{ext} C_0
+ L_2 S_0 - \frac{R^2 S_0^2 \theta_0}{L_2} + L_2  C_0 \theta_0
\right\} \theta
\nonumber \\ & \equiv &
\theta /\alpha^\mathrm{ext}.
\end{eqnarray}
The proportionality constant is
$\alpha \equiv \mathrm{d}\theta /\mathrm{d}z_\mathrm{t}$.
It has a different value for each of the three cases
$\mu > 0$ (contact, extension),
$\mu < 0$ (contact, retraction),
and $\mu = 0$ (non-contact).

%%%%%%%%%%%%%%%%%%%%%%%%%%%%%%%%%%%%%%%%%%%%%%%%
\subsubsection{Linear Analysis of Measured Data}

The measured data in the atomic force microscope
consists of the raw photo-diode voltage $\tilde V(t)$
and the piezo-drive position $z_\mathrm{p}(t)$.
These are both a function of time,
and so the voltage may equivalently be regarded as a function of position,
$\tilde V(z_\mathrm{p})$.
One has in fact two sets of data,
one for extend, $\tilde V^\mathrm{ext}(z_\mathrm{p})$,
and one for retract, $\tilde V^\mathrm{ret}(z_\mathrm{p})$.
Only the equations for extension will be shown explicitly here.

The tilde on the voltage is used to denote the raw measured voltage.
The raw voltage contains contributions from the change in angle
of the cantilever and from various artifacts that include
a constant voltage off-set,
thermal drift,
and virtual deflection.\cite{Zhu11a,Zhu11b}
Two further physical effects have to be carefully accounted for,
namely the drag force on the cantilever
and the long range asymptote of the surface force.
The deflection angle of the cantilever due to the surface forces
is what is desired to extract from the measured voltage.
The notation $V(z_\mathrm{p})$ will be used to denote the measured voltage
that has been corrected for these various artifacts and forces.

In the base-line region, where the surfaces are far from contact,
the surface force is small and in many cases negligible.
Hence almost all of the measured voltage in this region
is due to the artifacts just mentioned.
These in general are linear functions of position,
and one can define the measured base-line voltage on extension as
\begin{equation} \label{Eq:tilde-V-b}
\tilde V^\mathrm{ext}_\mathrm{b}(z_\mathrm{p})
=
\tilde V^\mathrm{ext}_\mathrm{b}
+ \tilde \beta^\mathrm{ext}_\mathrm{b} [z_\mathrm{p} - z_\mathrm{pb} ],
\end{equation}
and similarly for retraction.
The base-line slope is $ \tilde \beta^\mathrm{ext}_\mathrm{b}
\equiv
\left.
\mathrm{d} \tilde V^\mathrm{ext}(z_\mathrm{p})/ \mathrm{d}z_\mathrm{p}
\right|_{z_\mathrm{pb} } $.
The coefficients for this are obtained by a linear fit
to the measured data in an interval about the fixed position
$z_\mathrm{pb} $ in the base-line region.
Once the coefficients are determined,
this linear fit is applied to the whole measured regime,
not just the base-line region
(because the artifacts that it removes apply to the whole regime).
With this the corrected voltage  on extension is
\begin{equation}
V^\mathrm{ext}(z_\mathrm{p})
=
\tilde V^\mathrm{ext}(z_\mathrm{p})
-
\tilde V^\mathrm{ext}_\mathrm{b}(z_\mathrm{p}) .
\end{equation}
This is zero in the base-line region.

This expression removes from the raw signal
not only the artifacts mentioned above
but also the constant drag force.
(In some cases the drag force is not constant.\cite{Zhu11a,Zhu11b}
This effect,
which can be important for cantilevers with a low spring constant,
is not included in the present analysis.)
It also removes
the linear extrapolation of the asymptote of surface force,
\begin{equation}
F^\mathrm{ext}_\mathrm{b}(z_\mathrm{p})
=
F^\mathrm{ext}_\mathrm{b}
+ F^\mathrm{ext'}_\mathrm{b} [z_\mathrm{p} - z_\mathrm{pb} ].
\end{equation}
Here the constant force is
$F^\mathrm{ext}_\mathrm{b} = F^\mathrm{ext}(h_\mathrm{b})$
and the derivative is $F^\mathrm{ext'}_\mathrm{b}
= \mathrm{d} F^\mathrm{ext}(h_\mathrm{b})/\mathrm{d} h_\mathrm{b}$,
where the separation is
$h_\mathrm{b} = z_\mathrm{pb} + z_0^\mathrm{ext}$.
(This neglects the deflection of the cantilever,
which should be negligible;
if not, add to the separation
$z _\mathrm{tb} \approx F^\mathrm{ext}_\mathrm{b}/k_0$.)
In almost all cases this extrapolated surface force
from the  base-line region is negligible.
In those cases where it isn't, it has to be added back,
as will be done shortly.

The contact region is where the separation between the surfaces is zero,
and the tip moves equal and opposite to the piezo-drive,
$\Delta z_\mathrm{t} = -\Delta z_\mathrm{p}$.\cite{NB1}
In the  linear regime, the slope is constant
and this is also called the constant compliance regime.
It does not matter whether one fits the raw data or the corrected
data because the two contact slopes are related by
\begin{equation} \label{Eq:beta-c}
\beta^\mathrm{ext}_\mathrm{c}
\equiv \frac{\Delta V^\mathrm{ext}_\mathrm{c}}{\Delta z_\mathrm{p}}
, \;\;
\tilde \beta^\mathrm{ext}_\mathrm{c}
\equiv \frac{\Delta \tilde V^\mathrm{ext}_\mathrm{c}}{\Delta z_\mathrm{p}}
, \;\;
\beta^\mathrm{ext}_\mathrm{c}
=
\tilde \beta^\mathrm{ext}_\mathrm{c}
-
\tilde \beta^\mathrm{ext}_\mathrm{b} .
\end{equation}
If a positive voltage corresponds to a repulsive force,
then the slope ought to be negative.

The light lever measures the angle of the cantilever.
The key to analyzing atomic force microscope force data
is to calibrate the light lever by measuring
the proportionality constant between angle  and photo-diode voltage,
\begin{equation} \label{Eq:dV/dq}
\gamma \equiv \frac{\Delta V}{\Delta \theta} .
\end{equation}
This is the same on extension and retraction,
and it is the same in contact and out of contact.
This expression assumes a linear photo-diode,
but not necessarily a linear cantilever.

The value of this conversion factor follows from
the measured slope in contact, Eq.~(\ref{Eq:beta-c}),
and the linear proportionality between angular deflection
and tip position, Eq.~(\ref{Eq:ztQ-lin}).
Evaluating these on extension in contact one has
\begin{equation} \label{Eq:gamma-c}
\gamma^\mathrm{ext} =
\frac{\Delta V}{\Delta z_\mathrm{p}}
\frac{\Delta z_\mathrm{p}}{\Delta z_\mathrm{t}}
\frac{\Delta z_\mathrm{t}}{\Delta \theta}
=
- \beta^\mathrm{ext}_\mathrm{c} / \alpha^\mathrm{ext} .
\end{equation}
One has a similar result for retraction in contact
$\gamma^\mathrm{ret} = - \beta^\mathrm{ret}_\mathrm{c} /\alpha^\mathrm{ret}$.
Since this has to be a property of the light lever,
the value of $\gamma$ cannot depend
upon whether or not the surfaces are in contact,
or whether the measurement is made on extension or on retraction.
Hence one must have
$ \gamma^\mathrm{ext} = \gamma^\mathrm{ret} $,
or
\begin{equation}
\frac{ \beta^\mathrm{ext}_\mathrm{c} }{ \beta^\mathrm{ret}_\mathrm{c} }
=
\frac{\alpha(\mu)}{\alpha(-\mu)} ,
\end{equation}
since $\alpha^\mathrm{ext} = \alpha(\mu) $
and $\alpha^\mathrm{ret} = \alpha(-\mu) $.
The left hand side is a measured quantity,
and the right hand side is a known non-linear function of $\mu$,
Eq.~(\ref{Eq:ztQ-lin}).
There exist sophisticated algorithms for solving such non-linear equations,
with perhaps the most common if not the most powerful
being to guess the solution.
(This can be turned into a quadratic equation for $\mu$
if one expands $D(\mu)$ to leading order in $L_2/L_0$.
There is a small loss of accuracy in such an expansion,
which is not compensated by the even smaller gain
of an explicit analytic solution.)
This result provides a way of measuring the friction coefficient.

With $\gamma$ having been obtained from the measured contact slope
and the calculated rate of change of tip position with angle,
Eq.~(\ref{Eq:gamma-c}),
one can now give the surface force as a function of separation.
From Eq.~(\ref{Eq:QFz-lin}), the angle deflection is
\begin{equation}
\theta^\mathrm{ext}(z_\mathrm{p})
=
\gamma^{-1} V^\mathrm{ext}(z_\mathrm{p})
%} && \nonumber \\ &&
+
%\left\{ \frac{L_0^2}{2B} \left[ C_0 + \mu S_0 \right]
%+ \frac{ L_0  L_2}{B}  \left[ S_0 - \mu C_0  \right] \right\}
E^\mathrm{nc}
F^\mathrm{ext}_\mathrm{b}(z_\mathrm{p}) .
\end{equation}
Here the contribution of the linear extrapolation
of the asymptote of the surface force,
$F^\mathrm{ext}_\mathrm{b}(z_\mathrm{p}) $,
which is assumed a known function
and which was removed from the raw voltage signal,
has been added back to give the full deformation angle.
Note that it is the non-contact value of the conversion factor,
$E^\mathrm{nc} = E(\mu=0)$ that is used here.
Inserting this into Eq.~(\ref{Eq:QFz-lin}) gives the surface force
$F^\mathrm{ext}(z_\mathrm{p})$.
Explicitly in terms of the voltage it is
\begin{equation}
F^\mathrm{ext}(z_\mathrm{p}) =
\left\{
\begin{array}{cc} \displaystyle
\frac{1}{ E^\mathrm{nc} \gamma}  V^\mathrm{ext}(z_\mathrm{p})
+
F^\mathrm{ext}_\mathrm{b}(z_\mathrm{p}) ,
& h > 0,  \\ \displaystyle
\frac{1}{ E^\mathrm{ext} \gamma}  V^\mathrm{ext}(z_\mathrm{p})
+
F^\mathrm{ext}_\mathrm{b}(z_\mathrm{p}),
& h = 0 .
\end{array} \right.
%\frac{2B \gamma^{-1} V^\mathrm{ext}(z_\mathrm{p})
%}{
%L_0^2 \left[ C_0 + \mu S_0 \right]
%+ 2 L_0  L_2 \left[ S_0 - \mu C_0  \right] }
%\nonumber \\ && \mbox{ }
%+ F^\mathrm{ext}_\mathrm{b}(z_\mathrm{p}) .
\end{equation}
From Eq.~(\ref{Eq:ztQ-lin}) the separation is
\begin{eqnarray}
h^\mathrm{ext}(z_\mathrm{p})
& = &
z_\mathrm{p}
+ \frac{1}{ \alpha^\mathrm{nc} } \theta^\mathrm{ext}(z_\mathrm{p})
+ z_0^\mathrm{ext}
\\ \nonumber & = &
z_\mathrm{p}
+ \frac{1}{ \alpha^\mathrm{nc} \gamma } V^\mathrm{ext}(z_\mathrm{p})
+ \frac{E^\mathrm{nc}}{ \alpha^\mathrm{nc} }
F^\mathrm{ext}_\mathrm{b}(z_\mathrm{p})
+ z_0^\mathrm{ext} .
\end{eqnarray}
The constant $z_0^\mathrm{ext}$ is calculated so that $h^\mathrm{ext} = 0$
when the surfaces first come into contact (see next).
The separation equation is normally used explicitly for $h > 0$.
These three equations are written explicitly for extension;
for retraction in contact change the superscript `ext' to `ret',
including $E^\mathrm{ext} \equiv E(\mu)
\Rightarrow E^\mathrm{ret} \equiv E(-\mu)$.

%%%%%%%%%%%%%%%%%%%%%%%%%%%%%%%%%%%
\subsubsection{Zero of Separation} \label{Sec:z0-lin}

The constant $z_0^\mathrm{ext} $ remains to be determined.
The conventional way of establishing the zero of separation is by eye,
which is to say the force curve is shifted horizontally
until it looks `right'.
The problem with this is that there is often ambiguities
in identifying first contact, particularly when one has a steeply repulsive
surface force prior to contact.
Also the constant that gives $h=0$ at first contact may be
different to the constant that gives $h=0$ for most of the contact region,
even when the correct friction coefficient is used,
as will be demonstrated by explicit data below.
Finally, choosing the zero of separation by eye
introduces a psychological element into the analysis
and the potential for personal bias
that would be best removed by having a
mathematical algorithm for finding contact.

It may be objected that identifying the contact and the base-line regions
already introduce some form of psychological bias into
the analysis of the experimental data.
However, it turns out that the various fits
are not very sensitive to the choice of the region used for the fit,
within reason,
and the result do not vary significantly with different choices.
The zero of separation, however, feeds directly into the final result,
and a difference of as small as 1$\,$nm can
quantitatively effect the values of parameters that one is trying to measure
(e.g.\ the slip length in drainage flow can be of the same order),
and it can even qualitatively effect the physical
interpretation of the data.

The strategy is to define $z_0^\mathrm{ext} $
so that the separation is exactly zero at first contact,
(and to define $z_0^\mathrm{ret} $ so that $h=0$ at last contact).
`First' (or `last') contact is defined to mean the point at which
the extrapolated base-line voltage intersects the extrapolated contact voltage.
This definition is precise and unambiguous,
it is able to be calculated  mathematically,
and it is physically reasonable and in accord with one intuitive
understanding of the meaning of contact.

It should be understood that there are conceptual problems
with the meaning of `separation' at the molecular level.
The separation  as defined here,
$h \equiv z_\mathrm{p} + z_\mathrm{t} + z_0$,
is not precisely zero over the whole contact region.
It rather measures the difference between changes in the piezo-drive position
and changes in the tip position.
It is positive when there is a protuberance on the substrate,
and it is negative when there is a depression.
It is also negative when compression of a deformable surface occurs.
Hence the separation $h$ in contact really gives a topographic map
of the substrate.
The zero plane of the map is here defined
as the plane passing through the point of first or last contact.

Let $z_\mathrm{pc}^\mathrm{ext}$
be the piezo-drive position at first contact,
and let $V_\mathrm{c}^\mathrm{ext}
= V^\mathrm{ext}(z_\mathrm{pc}^\mathrm{ext})$ be the
corrected voltage at first contact.
(In the linear case,
it makes no difference to the results
what point is selected for  $z_\mathrm{pc}^\mathrm{ext}$.)
In contact,
$ V^\mathrm{ext}(z_\mathrm{p}) = V_\mathrm{c}^\mathrm{ext}
+ \beta_\mathrm{c}^\mathrm{ext} [z_\mathrm{p}- z_\mathrm{pc}^\mathrm{ext}]$.
The position at which the voltage in contact extrapolates to zero,
which is defined as first contact, is
\begin{equation}
z_\mathrm{pcb}^\mathrm{ext} =
z_\mathrm{pc}^\mathrm{ext}
- V_\mathrm{c}^\mathrm{ext}/\beta_\mathrm{c}^\mathrm{ext} .
\end{equation}
(The base-line corrected voltage is zero,
and so this is the same as the intersection of the
base-line and contact raw voltages.)
When the voltage is zero the angular deflection is
\begin{equation}
\theta^\mathrm{ext}(z_\mathrm{pcb}^\mathrm{ext})
=
E^\mathrm{nc}
F^\mathrm{ext}_\mathrm{b}(z_\mathrm{pcb}^\mathrm{ext}) .
\end{equation}
Inserting this into the equation for the separation,
and setting the latter to zero,
$ h^\mathrm{ext}(z_\mathrm{pcb}^\mathrm{ext}) = 0$,
gives the shift constant,
\begin{equation}
z_0^\mathrm{ext}
=
- z_\mathrm{pcb}^\mathrm{ext}
- \frac{E^\mathrm{nc}}{ \alpha^\mathrm{nc} }
F^\mathrm{ext}_\mathrm{b}(z_\mathrm{pcb}^\mathrm{ext}) .
\end{equation}

%%%%%%%%%%%%%%%%%%%%%%%%%%%%%%%%%%%%%%%%%%%%
\subsubsection{Effective Spring Constant} \label{Sec:keff}

The conventional modeling of the atomic force microscope
is not only linear but also effectively takes the cantilever
to be horizontal.
Ignoring the tilt is equivalent to equating
the cantilever deflection
to the vertical position of the tip, $x \equiv z_\mathrm{t}$.
The relationship between the measured photo-diode voltage
and the vertical tip position is given by the calibration
factor obtained from the slope of the contact region.
In this case,
the effective spring constant
that gives the non-contact force is
\begin{equation} \label{Eq:keff}
k_\mathrm{eff} \equiv
\frac{F_z}{z_\mathrm{t}}
= \alpha^\mathrm{nc} / E^\mathrm{nc} .
\end{equation}
The constants
$E^\mathrm{nc} \equiv E(\mu=0)$ and
$\alpha^\mathrm{nc} \equiv  \alpha(\mu=0)$
are defined in Eqs~(\ref{Eq:QFz-lin}) and (\ref{Eq:ztQ-lin}),
respectively.

The difference between the cantilever spring constant $k_0 = 3B/L_0^3$
and the effective spring constant $k_\mathrm{eff}$ can be substantial.
For the case analyzed in detail below
($L_0 = 110\,\mu$m, $R = 10.1\,\mu$m, $\theta_0 = -11^\circ$),
the cantilever spring constant is $k_0 = 1.37\,$N/m
and the effective spring constant is $k_\mathrm{eff} = 1.68\,$N/m.

In using the equations for the tilted cantilever to
convert measured atomic force microscope data to force,
one should use the cantilever spring constant.
In using the equations for the horizontal cantilever
(simple spring model, the conventional approach)
to convert measured atomic force microscope data to force,
one should use the effective spring constant.
In calculating a theoretical force curve
modeled with the cantilever as a simple spring,
one should also use  the effective spring constant.

Finally,
in Ref.~\onlinecite{Attard06a}
the correct equations for the thermal calibration
of the atomic force microscope cantilever were given.
In that paper the cantilever spring constant was denoted $k_0$
(here also denoted $k_0$),
and the effective force measuring spring constant
was denoted $k$ (here denoted $k_\mathrm{eff}$)
and was given in terms of the cantilever spring constant
in Eq.~(17) of Ref.~\onlinecite{Attard06a}.
In the present notation,
the correct thermal calibration method gives
the cantilever spring constant as
\begin{eqnarray} \label{Eq:k-therm}
k_0
%^{1/2}
& = &
\left\{
\frac{-2 \beta_\mathrm{cb} L_0 }{3[ C_0 + 2 L_2 S_0 /L_0 ] }
\right.  \\ \nonumber && \mbox{ } \times \left.
0.7830 \sqrt{\frac{6 k_\mathrm{B} T}{\pi L_0^2 f_\mathrm{R} P_\mathrm{DC} Q}}
\right. \\ \nonumber && \mbox{ } \times \left.
\left[ C_0^2 + ( 3 L_2 S_0 C_0/L_0 )
+ 3 S_0^2 L_2^2 /L_0^2 \right]
 \rule{0cm}{.5cm} \right\}^2. % strut:
\end{eqnarray}
Here the measured quantities are
$\beta_\mathrm{cb} = \Delta V /\Delta z_\mathrm{p}$,
which is the contact slope evaluated near the base-line voltage,
(the average of the extend and retract values),
$f_\mathrm{R}$,
which is the resonance frequency of the first mode in Hz,
$P_\mathrm{DC}$,
which  is the direct current power response in V$^{2}\,$Hz$^{-1}$,
and $Q$, which is the quality factor. %cf Ref.\0nlinecite{Higgins06}
The length of the rigid part at the end of the cantilever, $L_1$,
defined in earlier analyses\cite{Attard98,Attard99,Attard06a}
has here and throughout been set to zero.

This inserted into Eq.~(\ref{Eq:keff}) gives the effective
spring constant for use when the cantilever is modeled as a
simple spring (i.e.\ tilt neglected),
which is usually the case in the linear analysis of measured data
and the theoretical modeling of force-separation curves.

%%%%%%%%%%%%%%%%%%%%%%%%%%%%%%%%%%%%%%%%%%%%
\subsubsection{Effective Drag Length} \label{Sec:Leff}

The above procedure for analyzing the measured data
removes the constant force due to the drag on the cantilever
from the extension data,
and its equal and opposite value from the retraction data.
In some case it is useful to have available an explicit value
for this drag force.

Like the drainage force,
and unlike the virtual deflection,
the contribution to the gradient of the base-line due to thermal drift
is equal and opposite on extension and retraction.

With $\tilde \beta_\mathrm{b}^\mathrm{ext}$ and
$\tilde \beta_\mathrm{b}^\mathrm{ret}$
being the measured base-line slopes of the raw voltage as defined above,
and
$F_\mathrm{b}^\mathrm{ext'}
=-F_\mathrm{b}^\mathrm{ret'}$
being the gradient of the drainage force in the base-line region,
then the gradient of the voltage due to thermal drift is
\begin{equation}
\frac{\mathrm{d} \tilde V^\mathrm{ext}_\mathrm{th} }{ \mathrm{d} z_\mathrm{p} }
=
\frac{1}{2} \left[
\tilde \beta_\mathrm{b}^\mathrm{ext}- \tilde \beta_\mathrm{b}^\mathrm{ret}
\right]
-
\frac{\gamma_\mathrm{b} E^\mathrm{nc} }{2} \left[
F_\mathrm{b}^\mathrm{ext'} - F_\mathrm{b}^\mathrm{ret'} \right] .
\end{equation}
This is equal and opposite to the gradient on retraction.
With $z_\mathrm{p,turn}$ being the turn point at the end of the extend
branch and the beginning of the retract branch,
it is readily shown that the constant drag force on extension is
\begin{eqnarray}
F_\mathrm{drag}^\mathrm{ext}
& = &
-F_\mathrm{b}^\mathrm{ext}
+
\frac{1}{2\gamma E^\mathrm{nc}}
\left[ \tilde V_\mathrm{b}^\mathrm{ext} - \tilde V_\mathrm{b}^\mathrm{ret}
\right]
 \\ \nonumber && \mbox{ }
-
\frac{1}{2\gamma E^\mathrm{nc}}
\frac{\mathrm{d} \tilde V^\mathrm{ext}_\mathrm{th} }{ \mathrm{d} z_\mathrm{p} }
\left[ 2 z_\mathrm{p,turn} - z_\mathrm{pb}^\mathrm{ext}
- z_\mathrm{pb}^\mathrm{ret} \right] .
\end{eqnarray}
One can define an effective drag length from
\begin{equation}
F_\mathrm{drag}^\mathrm{ext}
\equiv
- 6 \pi \eta \dot z_\mathrm{p}^\mathrm{\,ext} L_\mathrm{eff} .
\end{equation}
One should not take $L_\mathrm{eff} $ too literally,
but it is expected to be somewhat less than the length of the cantilever,
typically one third to one half of $L_0$.
It should be independent of the drive velocity,
although because it is derived from the difference in the base-line values,
it can have relative large errors,
on the order of 10--20\% (see results below).

%\newpage
%%%%%%%%%%%%%%%%%%%%%%%%%%%%%%%%%%%%%%%%%%%%%%%%%%%%%%%%%%%%%%%%%%%%%%
%
\section{Non-Linear Cantilever or Photo-diode} \label{Sec:non-linear}
%
%%%%%%%%%%%%%%%%%%%%%%%%%%%%%%%%%%%%%%%%%%%%%%%%%%%%%%%%%%%%%%%%%%%%%%

%%%%%%%%%%%%%%%%%%%%%%%%%%%%%%%%%%%%%%%%%%%%%%%%
\subsection{Non-Linear  Cantilever} \label{Sec:nl-cant}

For a spherical colloid probe of radius $R$,
simple geometry, Fig.~\ref{Fig:AFMgeom},
gives the lever arm  as
\begin{equation} \label{Eq:L2Q}
L_2(\theta_\mathrm{tot} ) =
R \sqrt{ 2 +  2\cos(  \theta_0 + \theta ) } .
\end{equation}
In the undeflected state this will be written
$L_2 \equiv L_2(\theta_0 ) = R \sqrt{ 2 +  2\cos \theta_0  } $.
For the case of a tipped cantilever,
$L_2(\theta_\mathrm{tot} )$ is equal to the length of the tip,
and there is no dependence on the deflection angle.
In fact, even for a spherical probe
the dependence on angle is practically negligible,
and $L_2(\theta_\mathrm{tot} )$ can be replaced by $L_2$,
or even by $2R$.

The vertical position of the tip depends upon the deflection
and the deflection angle.
Again simple geometry shows that
\begin{equation} \label{Eq:ztQ}
z_\mathrm{t} =
x \cos \theta_\mathrm{tot}
+ L_2(\theta_\mathrm{tot}) \sin (\theta_\mathrm{tot}) \theta_\mathrm{tot}
- L_2 S_0 \theta_0 .
\end{equation}
Since it is the change in tip position that is important,
this has been chosen to be zero in the non-deflected state.
This equation is the major source of non-linearity
in the cantilever.

The force $F$ and torque $\tau$ on the cantilever,
which were treated in \S\ref{Sec:lin-horiz-mu=0},
are a linearly proportional to the normal force $F_z$
and to the lateral force $F_y$,
with the proportionality constant depending upon
the lever arm and the tilt angle,
\begin{equation}
F = F_z \cos \theta_\mathrm{tot}  + F_y \sin \theta_\mathrm{tot} ,
\end{equation}
and
\begin{equation}
\tau =
F_z L_2(\theta_\mathrm{tot}) \sin \theta_\mathrm{tot}
- F_y L_2(\theta_\mathrm{tot}) \cos \theta_\mathrm{tot} .
\end{equation}

Using the linear model of friction, $F_y = \pm \mu F_z $,
\S\ref{Sec:Friction},
the force and torque are linearly proportional
to the surface force $ F_z$.
For extension in contact one has
\begin{equation}
F =
\left[ \cos \theta_\mathrm{tot}  + \mu \sin \theta_\mathrm{tot} \right] F_z  ,
\end{equation}
and
\begin{equation}
\tau =
L_2(\theta_\mathrm{tot}) \left[ \sin \theta_\mathrm{tot}
- \mu \cos \theta_\mathrm{tot} \right] F_z .
\end{equation}
(Of course for retraction in contact, $\mu \Rightarrow -\mu$,
and out of contact, $\mu = 0$.)
Inserting these into the standard cantilever equations
(\ref{Eq:x(F,tau)}) and (\ref{Eq:q(F,tau)})
gives the deflection and deflection angle
as linearly proportional to the normal force
(on extension in contact),
\begin{eqnarray} \label{Eq:xFz}
x & = &
\left\{
\frac{L_0^3}{3B}
\left[ \cos \theta_\mathrm{tot}  + \mu \sin \theta_\mathrm{tot} \right]
\right. \nonumber \\ & & \left. \mbox{ } +
\frac{ L_0^2  L_2(\theta_\mathrm{tot}) }{2B}
\left[ \sin \theta_\mathrm{tot}
- \mu \cos \theta_\mathrm{tot} \right] \right\} F_z ,
\end{eqnarray}
and
\begin{eqnarray} \label{Eq:QFz}
\theta & = &
\left\{
\frac{L_0^2}{2B}
\left[ \cos \theta_\mathrm{tot}  + \mu \sin \theta_\mathrm{tot} \right]
\right. \nonumber \\ & & \left. \mbox{ } +
\frac{ L_0  L_2(\theta_\mathrm{tot}) }{B}
\left[ \sin \theta_\mathrm{tot}
- \mu \cos \theta_\mathrm{tot} \right] \right\} F_z .
\end{eqnarray}
Because the total angle depends upon the deflection angle,
$\theta_\mathrm{tot} = \theta_0 + \theta $,
these represent a non-linear relationship
between the force, deflection, and deflection angle.
This last equation is best written by taking the proportionality
function over to the other side,
which gives the force as an explicit non-linear function
of the deflection angle, $F_z(\theta)$.

%%%%%%%%%%%%%%%%%%%%%%%%%%%%%%%%%%%%%%%%%%%%%%%%
\subsection{Non-Linear Analysis of Measured Data} \label{Sec:nl-anal}

The non-linear analysis in this section holds
for both sources of non-linearity:
the cantilever treated explicitly in the preceding subsection
and the photo-diode non-linearity, for which no specific model is given.
In later sub-sections, one or other of these will be turned off.

The raw measured voltage $\tilde V$
is a non-linear function of the effective total angle,
$\tilde \theta_\mathrm{tot} =
\theta_\mathrm{tot} + \theta_\mathrm{b} - \theta_\mathrm{bf}
=
\theta_0 + \theta
+ \theta_\mathrm{b} - \theta_\mathrm{bf}$.
Note the distinction between the physical or relevant
total angle $ \theta_\mathrm{tot}$
and the effective total angle $\tilde \theta_\mathrm{tot} $.
The physical contributions are the  tilt angle $\theta_0$,
which is known,
and the deflection angle due to the surface force,
$\theta$,
which is to be obtained.
The base-line angle $\theta_\mathrm{b}$
is essentially an artifact arising from thermal drift
and virtual deflection,
and it includes the angle due to drag force,
here assumed constant
(but see Refs.~\onlinecite{Zhu11a,Zhu11b},
where variable drag  is shown to occur for weak cantilevers),
and the angle deflection
due to the linearly extrapolated asymptote of the surface force,
$\theta_\mathrm{bf}$.
In the non-linear case,
whether it be the non-linear cantilever or the non-linear photo-diode,
one cannot just subtract the base-line voltage from the raw voltage
to obtain a corrected voltage that gives $\theta$ directly.
However, the total angle is a linear function its component parts,
and so the immediate tasks are to obtain the effective total angle
from the measured raw voltage, $\tilde \theta_\mathrm{tot}(\tilde V)$,
and to obtain the base-line angle as a function of the piezo-drive position,
$\theta_\mathrm{b}(z_\mathrm{p})$.

Do a non-linear fit of the measured photo-diode voltage
in the contact region on extension,
\begin{equation} \label{Eq:zp(V)}
z_\mathrm{pc}^\mathrm{ext}(\tilde V)
=
a_0^\mathrm{ext}
+ a_1^\mathrm{ext} \tilde V
+ a_2^\mathrm{ext} \tilde V^2
+ a_3^\mathrm{ext} \tilde V^3
+ \ldots
\end{equation}
It is best not to use too many terms in the fit.
Also one should be certain that the fit begins just after first contact,
and that it ends before any anomalies associated with the turn around point
at the end of the extension branch.

In the non-linear cantilever case,
one regards the angular deflection $\theta$ as the independent variable.
%(This $\theta$ is the same as the $\theta_\mathrm{f}$ used above.)
From the non-linear Eq.~(\ref{Eq:QFz}),
one can calculate the surface force $F_z(\theta)$,
and inserting this into the non-linear Eq.~(\ref{Eq:xFz})
one can calculate the deflection $x(\theta)$.
Inserting these into the non-linear equation Eq.~(\ref{Eq:ztQ})
one can calculate the tip position $z_\mathrm{t}(\theta)$.
Hence from these non-linear cantilever equations,
$z_\mathrm{t}(\theta;\mu)$ is easily calculated.
A non-linear fit can be made to this,
(in contact and on extension, $\mu > 0$),
\begin{equation} \label{Eq:q(zt)}
\theta^\mathrm{ext}_\mathrm{c}(z_\mathrm{t})
=
b_1^\mathrm{ext} z_\mathrm{t}
+ b_2^\mathrm{ext} z_\mathrm{t}^2
+ b_3^\mathrm{ext} z_\mathrm{t}^3
+\ldots
\end{equation}
The leading coefficient is
$b_1^\mathrm{ext} = \alpha^\mathrm{ext}$,
which was given analytically in Eq.~(\ref{Eq:ztQ-lin}).
One has $\theta^\mathrm{ext}_\mathrm{tot,c}(z_\mathrm{t})
= \theta_0 + \theta^\mathrm{ext}_\mathrm{c}(z_\mathrm{t})$.

%%%%%%%%%%%%%%%%

As in the linear case,
the measured voltage is fitted to a straight line
in the base-line region, Eq.~(\ref{Eq:tilde-V-b}),
\begin{equation}
\tilde V^\mathrm{ext}_\mathrm{b}(z_\mathrm{p})
=
\tilde V^\mathrm{ext}_\mathrm{b}
+ \tilde \beta^\mathrm{ext}_\mathrm{b} [z_\mathrm{p} - z_\mathrm{pb} ].
\end{equation}

Now the raw contact slope from the non-linear fit Eq.~(\ref{Eq:zp(V)}),
evaluated at the base-line voltage constant,
$\tilde V^\mathrm{ext}_\mathrm{b}$
(possibly an extrapolation beyond the region of the fit),
is
\begin{eqnarray} \label{Eq:beta-cb}
\tilde \beta_\mathrm{cb}^\mathrm{ext}
&\equiv &
\left. \frac{ \Delta \tilde V_\mathrm{c}^\mathrm{ext} }{ \Delta z_\mathrm{p} }
\right|_{\tilde V_\mathrm{b}^\mathrm{ext}}
\nonumber \\ & = &
\frac{1}{
a_1^\mathrm{ext}
+ 2 a_2^\mathrm{ext}
\tilde V^\mathrm{ext}_\mathrm{b}
+ 3 a_3^\mathrm{ext}
(\tilde V^\mathrm{ext}_\mathrm{b})^2
+ \ldots }  .
\end{eqnarray}
%This gradient is the sum of that due to the surface forces
%plus that due to the base-line,
%$\tilde \beta_\mathrm{cb}^\mathrm{ext}
%= \beta_\mathrm{cbf}^\mathrm{ext}
%+ \tilde \beta^\mathrm{ext}_\mathrm{b}
%- \beta^\mathrm{ext}_\mathrm{bf} $.
%The subscript `cbf' indicates the gradient in contact, `c',
%evaluated at the base-line voltage $\tilde V^\mathrm{ext}_\mathrm{b}$ ,`b',
%due to the surface force, `f'.
%Hence at this stage one has measured the rate of change of photo-diode voltage
%due solely to the change of drive position in contact,
%$\beta_\mathrm{cbf}^\mathrm{ext} =
%\tilde \beta_\mathrm{cb}^\mathrm{ext} -  \tilde \beta^\mathrm{ext}_\mathrm{b}
%+ \beta^\mathrm{ext}_\mathrm{bf}$.
%(Actually the surface force asymptote $\beta^\mathrm{ext}_\mathrm{bf}$
%remains to be obtained.)

The tip position is a linear function
of the deflection angle  in the base-line region,
and the gradient is
\begin{eqnarray} \label{Eq:alpha-bnc}
\alpha_\mathrm{b}^\mathrm{nc}
& \equiv &
\left. \frac{\mathrm{d}\theta}{\mathrm{d} z_\mathrm{t}}
\right|_{z_\mathrm{t}=0}
\\ \nonumber & = &
\frac{1}{
D^\mathrm{nc} C_0 + L_2 S_0 + L_2 C_0 \theta_0 - {R^2 S_0^2 \theta_0}/{L_2}
} ,
\end{eqnarray}
where
$D^\mathrm{nc} = D(0) = [2 L_0^3 C_0 + 3 L_0^2 L_2 S_0 ]
/[3 L_0^2 C_0 + 6 L_0 L_2 S_0 ]$,
as follow from Eqs (\ref{Eq:x(Q)}) and (\ref{Eq:ztQ-lin}).

The raw voltage is the same function of the total angle
in and out of contact,
and on extension and on retraction.
Hence the rate of change of raw voltage
with total angle (equivalently, deflection angle)
in the base-line region,
$\gamma_\mathrm{b}^\mathrm{ext}$,
can be evaluated in contact on extension
at the base-line voltage.
One has
\begin{eqnarray} \label{Eq:gamma_b}
\gamma_\mathrm{b}^\mathrm{ext}
& \equiv &
\left.
\frac{\Delta \tilde V}{\Delta \theta }
\right|_\mathrm{cb,ext} %,\tilde V_\mathrm{b}^\mathrm{ext}}
\\ \nonumber & = &
\left. \frac{\Delta z_\mathrm{t} }{\Delta \theta } \right|_\mathrm{cb,ext}
\left. \frac{\Delta z_\mathrm{p} }{\Delta z_\mathrm{t} } \right|_\mathrm{cb,ext}
\left. \frac{\Delta \tilde V}{\Delta z_\mathrm{p} } \right|_\mathrm{cb,ext}
\nonumber \\ & = &
- \tilde \beta_\mathrm{cb}^\mathrm{ext} /\alpha_\mathrm{b}^\mathrm{ext} .
\end{eqnarray}
Here $\alpha_\mathrm{b}^\mathrm{ext} \equiv \alpha_\mathrm{b}(\mu)$
is given by Eq.~(\ref{Eq:ztQ-lin}) with $\mu > 0$.
One can similarly obtain $\gamma_\mathrm{b}^\mathrm{ret}$.
The correct value of $\mu$ makes
$\gamma_\mathrm{b}^\mathrm{ext} = \gamma_\mathrm{b}^\mathrm{ret}$.
Actually,
the correct $\mu$ should make
$\Delta \tilde V/\Delta \theta $ equal on extend and retract
for all voltages.

With this conversion factor,
the angle corresponding to the base-line voltage is
\begin{equation} \label{Eq:Qb(zp)}
\theta_\mathrm{b}^\mathrm{ext}(z_\mathrm{p})  =
E^\mathrm{nc} F^\mathrm{ext}_\mathrm{b} +
(\gamma_\mathrm{b}^\mathrm{ext})^{-1}
\tilde \beta_\mathrm{b}^\mathrm{ext}
[ z_\mathrm{p} - z_\mathrm{pb}^\mathrm{ext} ] .
\end{equation}
This makes the base-line angle  at $z_\mathrm{pb}^\mathrm{ext}$
equal to that due to the surface force alone;
$E^\mathrm{nc} = E(0)$, Eq.~(\ref{Eq:QFz-lin}),
converts force to angle in the linear regime of low force.
The change in angle from its value at $z_\mathrm{b}^\mathrm{ext}$,
includes the contribution from the linear extrapolation
of the asymptote of the surface force.
The latter is
\begin{equation} \label{Eq:Qbf(zp)}
\theta_\mathrm{bf}^\mathrm{ext}(z_\mathrm{p})  =
E^\mathrm{nc}
\left\{ F^\mathrm{ext}_\mathrm{b}
+ F^\mathrm{ext'}_\mathrm{b} [z_\mathrm{p} - z_\mathrm{pb}^\mathrm{ext}]
\right\} .
%^\mathrm{ext}_\mathrm{b}(z_\mathrm{p}) .
\end{equation}
These last two equations apply for all $z_\mathrm{p}$,
not just in the base-line region.
The constant factors do not change value in or out of contact.
One can see that $\theta_\mathrm{b}^\mathrm{ext}(z_\mathrm{p})
- \theta_\mathrm{bf}^\mathrm{ext}(z_\mathrm{p}) $
is the virtual angle without any contribution from the surface force asymptote.
Hence
$\tilde \theta_\mathrm{tot} = \theta_0 + \theta -
\theta_\mathrm{b}^\mathrm{ext}(z_\mathrm{p})
- \theta_\mathrm{bf}^\mathrm{ext}(z_\mathrm{p})$
is the total effective angle (physical plus virtual)
without double counting the surface force.

From Eq.~(\ref{Eq:zp(V)}),
$z_\mathrm{pc}(\tilde V)$ is known in contact.
The tip position is related to the piezo-drive position in contact by
$0 = h \equiv z_\mathrm{pc} + z_\mathrm{tc} + z_0^\mathrm{ext}$, or
\begin{equation} \label{Eq:ztc(V)}
z_\mathrm{tc}(\tilde V) = z_0^\mathrm{ext} - z_\mathrm{pc}(\tilde V) .
\end{equation}
(Strictly, this holds at one particular position in contact.
See the discussion in the third and fourth paragraphs
of \S\ref{Sec:z0-lin}.)
The constant $z_0^\mathrm{ext} $ will be determined explicitly
in the following subsection;
in the mean time it can be regarded as an arbitrary horizontal shift
that establishes contact at zero separation.
Using successively Eqs~(\ref{Eq:zp(V)}), (\ref{Eq:ztc(V)}),
and (\ref{Eq:q(zt)}),
one can calculate the deflection angle for any given raw voltage in contact,
$\theta_\mathrm{c}(\tilde V)$.
In contact,
the total effective angle as a function of raw voltage is
\begin{equation}
\tilde \theta_\mathrm{tot,c}^\mathrm{ext}(\tilde V)
=
\theta_0 + \theta_\mathrm{c}^\mathrm{ext}(\tilde V)
+ \theta_\mathrm{b}^\mathrm{ext}(z_\mathrm{pc}(\tilde V))
- \theta_\mathrm{bf}^\mathrm{ext}(z_\mathrm{pc}(\tilde V)) .
\end{equation}
This is the desired relationship
between the raw photo-diode voltage and the total cantilever angle.
It is most convenient to do a non-linear fit of this
(necessarily for voltages in the contact region)
\begin{equation} \label{Eq:Qtotc}
\tilde \theta_\mathrm{tot,c}^\mathrm{ext}(\tilde V)
=
c_0^\mathrm{ext} + c_1^\mathrm{ext} \tilde V + c_2^\mathrm{ext} \tilde V^2
+ \ldots
\end{equation}
Although this is explicitly derived for contact on extension,
it ought equal the analogous result  for contact on retraction,
(assuming the correct value of $\mu$,
and the validity of the linear friction law).
Either expression can be used unchanged for any $\tilde V$
measured out of contact and on extension or retraction.
Hence the left hand side will be written
simply $\tilde \theta_\mathrm{tot}(\tilde V)$.

At a given measured datum  $\{z_\mathrm{p},\tilde V\}$,
in or out of contact, on extension,
the deflection angle is
\begin{equation} \label{Eq:Q(zp,V)}
\theta(z_\mathrm{p},\tilde V) =
\tilde \theta_\mathrm{tot}(\tilde V) - \theta_0
- \theta_\mathrm{b}^\mathrm{ext}(z_\mathrm{p})
+ \theta_\mathrm{bf}^\mathrm{ext}(z_\mathrm{p}) .
\end{equation}
The total physical angle is of course
$\theta_\mathrm{tot}(z_\mathrm{p},\tilde V)
= \theta_0 + \theta(z_\mathrm{p},\tilde V) $.
From this the force $F(z_\mathrm{p},\tilde V)$
follows from Eq.~(\ref{Eq:QFz}),
and the separation
$h(z_\mathrm{p},\tilde V)
= z_\mathrm{p} + z_\mathrm{t}(z_\mathrm{p},\tilde V) + z_0^\mathrm{ext}$
follows from  Eqs~(\ref{Eq:ztQ}) and (\ref{Eq:xFz}).

Note that in the non-linear analysis,
this equation for the separation is applied both in and out of contact;
the separation is not simply set to zero in contact
(c.f.~Fig.~\ref{Fig:VvsZp}).
Non-zero values of $h$ in the contact region
give information about the topography
and other physical attributes of the surfaces,
as will be shown below.

%%%%%%%%%%%%%%%%%%%%%%%%%%%%%%%%
\subsubsection{Zero of Separation} \label{Sec:zero-sep-nl}

The constant $z_0^\mathrm{ext} $ remains to be determined
in the non-linear case.
As discussed in the linear case in \S\ref{Sec:z0-lin} above,
there are good reasons for desiring a mathematical algorithm
for determining the zero of separation
that avoids guess-work or personal bias.
As in the linear case,
one defines the plane of zero separation
as passing through the point of first (or last contact),
and one defines  the point of first (or last contact)
as the point of intersection of the extrapolations
of the contact and of the base-line raw voltages.
In the non-linear case this is accomplished  as follows.

Begin by defining $z_\mathrm{pcb}^\mathrm{ext}$
as the piezo-drive position at which
the fitted voltage in contact,
Eq.~(\ref{Eq:zp(V)}),
 equals the constant part of the base-line voltage,
 $\tilde V^\mathrm{ext}_\mathrm{b}$,
\begin{eqnarray}
z_\mathrm{pcb}^\mathrm{ext}
& = &
a_0^\mathrm{ext}
+ a_1^\mathrm{ext}
\tilde V^\mathrm{ext}_\mathrm{b}
+ a_2^\mathrm{ext}
[\tilde V^\mathrm{ext}_\mathrm{b}]^2
%\nonumber \\ && \mbox{ }
%+ a_3^\mathrm{ext}
%[\tilde V^\mathrm{ext}_\mathrm{b}]^3
+ \ldots
\end{eqnarray}
Now a small correction to this will be made such that
$z_\mathrm{pcb}^\mathrm{ext,*}$ is the piezo-drive position at which
the fitted voltage in contact equals
the actual base-line voltage at that position,
\emph{viz}. $z_\mathrm{pcb}^\mathrm{ext,*}
= z_\mathrm{pc}^\mathrm{ext}(\tilde V_\mathrm{b}^\mathrm{ext}
(z_\mathrm{pcb}^\mathrm{ext,*}))$.
Using a Taylor expansion to linear order
about $\tilde V_\mathrm{b}^\mathrm{ext}$,
one has
\begin{eqnarray}
z_\mathrm{pcb}^\mathrm{ext,*} - z_\mathrm{pcb}^\mathrm{ext}
& = &
\frac{\mathrm{d} z_\mathrm{p} }{\mathrm{d}  \tilde V^\mathrm{ext}_\mathrm{c} }
\left[ \tilde V_\mathrm{b}^\mathrm{ext}(z_\mathrm{pcb}^\mathrm{ext,*})
- \tilde V_\mathrm{b}^\mathrm{ext} \right]
\nonumber \\ & = &
(\tilde \beta_\mathrm{cb}^\mathrm{ext})^{-1}
\tilde \beta_\mathrm{b}^\mathrm{ext}
\left[ z_\mathrm{pcb}^\mathrm{ext,*}
- z_\mathrm{pb}^\mathrm{ext} \right] ,
\end{eqnarray}
or
\begin{equation} \label{Eq:zpcb*}
z_\mathrm{pcb}^\mathrm{ext,*}
=
\frac{\tilde \beta_\mathrm{cb}^\mathrm{ext} z_\mathrm{pcb}^\mathrm{ext}
- \tilde \beta_\mathrm{b}^\mathrm{ext} z_\mathrm{pb}^\mathrm{ext}
}{\tilde \beta_\mathrm{cb}^\mathrm{ext}
- \tilde \beta_\mathrm{b}^\mathrm{ext} } .
\end{equation}
Since in general $ \left|\tilde \beta_\mathrm{cb}^\mathrm{ext} \right|
\gg \left | \tilde \beta_\mathrm{b}^\mathrm{ext} \right| $,
the difference between
$z_\mathrm{pcb}^\mathrm{ext,*}$ and $z_\mathrm{pcb}^\mathrm{ext}$
is generally small, in many cases negligible.

At this particular position,
the voltages extrapolated from contact and from the base-line are equal,
$\tilde V^* \equiv
\tilde V_\mathrm{c}(z_\mathrm{pcb}^\mathrm{ext,*})
=
\tilde V_\mathrm{b}(z_\mathrm{pcb}^\mathrm{ext,*})$,
and hence the corresponding effective total  angles
must also be equal,
$\tilde \theta_\mathrm{tot,c}^\mathrm{ext}(\tilde V^*)
= \tilde \theta_\mathrm{tot,b}^\mathrm{ext}(\tilde V^*)$.
The left hand side is
\begin{equation}
\tilde \theta_\mathrm{tot,c}^\mathrm{ext}(\tilde V^*)
=
\theta_0
+ \theta_\mathrm{c}^\mathrm{ext}(\tilde V^*)
+ \theta_\mathrm{b}^\mathrm{ext}(z_\mathrm{pcb}^\mathrm{ext,*})
- \theta_\mathrm{bf}^\mathrm{ext}(z_\mathrm{pcb}^\mathrm{ext,*}) ,
\end{equation}
and the right hand side is
\begin{equation}
\tilde \theta_\mathrm{tot,b}^\mathrm{ext}(\tilde V^*)
=
\theta_0
+ \theta_\mathrm{b}^\mathrm{ext}(z_\mathrm{pcb}^\mathrm{ext,*}) .
\end{equation}
Hence at this particular position,
the deflection of the cantilever due to surface forces
is equal to the linearly extrapolated deflection due to
the base-line surface force,
\begin{equation} \label{Eq:qf(zpb)}
\theta_\mathrm{c}^\mathrm{ext}(\tilde V^*)
=
\theta_\mathrm{bf}^\mathrm{ext}(z_\mathrm{pcb}^\mathrm{ext,*}) ,
\end{equation}
with $\theta_\mathrm{bf}^\mathrm{ext}(z_\mathrm{p})$
being given by Eq.~(\ref{Eq:Qbf(zp)}).
In this linear part of the curve,
the position of the tip at this angle is
\begin{eqnarray} \label{Eq:ztcb}
z_\mathrm{tcb}^\mathrm{ext,*}
& = &
\theta_\mathrm{bf}^\mathrm{ext}(z_\mathrm{pcb}^\mathrm{ext,*})
/\alpha_\mathrm{b}^\mathrm{nc}
\nonumber \\ & = &
E^\mathrm{nc}
\left\{ F^\mathrm{ext}_\mathrm{b}
+ F^\mathrm{ext'}_\mathrm{b}
 [z_\mathrm{pcb}^\mathrm{ext,*} - z_\mathrm{pb}^\mathrm{ext}]
\right\}  /\alpha_\mathrm{b}^\mathrm{nc} ,
\end{eqnarray}
with $\alpha_\mathrm{b}^\mathrm{nc} \equiv \alpha(0)$
being given by Eq.~(\ref{Eq:alpha-bnc}),
which is just Eq.~(\ref{Eq:ztQ-lin}) with $\mu = 0$,

This particular point is first contact (on extension;
it is last contact on retraction),
and the separation is set to zero,  $h=0$.
This gives the off-set as
\begin{equation}
z_0^\mathrm{ext} = - z_\mathrm{pcb}^\mathrm{ext,*} - z_\mathrm{tcb}^\mathrm{ext,*}.
\end{equation}
This quantity $z_0^\mathrm{ext}$
is used in Eqs (\ref{Eq:ztc(V)}) to  (\ref{Eq:Q(zp,V)}),
which equations were not used in its derivation.
Any difference between $z_0^\mathrm{ext}$ and $  z_0^\mathrm{ret}$
mainly reflects the different sign of the friction force on
extension and retraction,
$z_\mathrm{t}(\tilde V,\mu) \ne z_\mathrm{t}(\tilde V,-\mu)$.
The two values bring
the non-zero values of the separation in contact on the two branches
more or less into coincidence,
as will be demonstrated with the data below.

%%%%%%%%%%%%%%%%%%%%%%%%%%%%%%%%%%%%%%%%%%%%%%%%%%%%%%%%%%%%
\subsection{Non-Linear Cantilever, Linear Photo-diode} \label{Sec:nlcant-linpd}

%%%%%%%%%%%%%%%%%%%%%%%%%%%%%%%%%%%%%%%%%%%%%%%%%%%%%%%%%%%%%%%%%%
\begin{figure}[t!]
\centerline{
\resizebox{8.5cm}{!}{ \includegraphics*{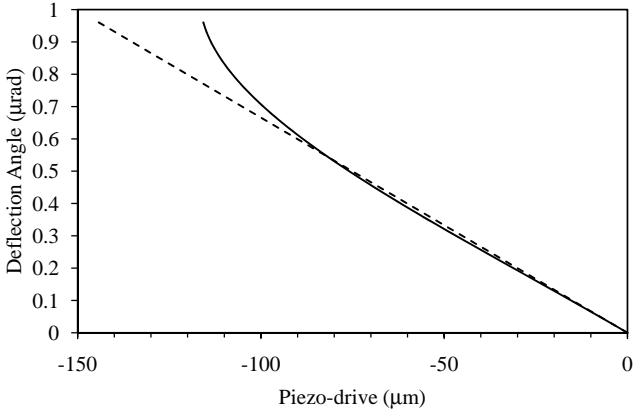} } }
% From My Documents\Papers\Current\nonlinear AFM\nlAFm-Nov12.xlsx:Chart1
\caption{\label{Fig:QvsZp-nlcant}
Calculated deflection angle versus the piezo-drive displacement
in contact on extension
($L_0=235\,\mu$m, $R=10.11\,\mu$m, $\theta_0=-11^\circ$,
and  $\mu = 0.375$).
The solid curve is the non-linear calculation,
and the dashed line is the linear calculation,
which gives the tangent at first contact.
Note that the total drive distance is rather large .
}
\end{figure}
%%%%%%%%%%%%%%%%%%%%%%%%%%%%%%%%%%%%%%%%%%%%%%%%%%%%%%%%%%%%%%%%%%

Figure \ref{Fig:QvsZp-nlcant}
shows the  deflection angle
versus piezo-drive position
calculated with the non-linear cantilever equations given above.
In the displayed data,
the photo-diode does not enter,
and so the figure gives an indication
of the non-linearity due to the cantilever alone.
The linear result,
the straight line,
is just the tangent to the curve  at first contact.
One can see that the initial non-linear effect
is for the angular deflection to increase at a slightly decreased rate
(i.e.\ it dips below the tangent;
the non-linear curve is initially concave down).
Subsequently the angular deflection increases at an increasing rate
and rises above the tangent,
(the non-linear curve becomes concave up).

It should be emphasized that the drive distances in contact
used in Fig.~\ref{Fig:QvsZp-nlcant},
although experimentally realizable,
are on the order of 100 times larger than are used in a typical
atomic force microscope force measurement.
The point of the figure is to show
what one might qualitatively expect from cantilever
non-linearities in an experiment
with large drive distances and a linear photo-diode.

%In the actual experiment shown in Fig.~\ref{Fig:VvsZp},
%the drive distance in  contact is about 300$\,$nm.
The non-linearity may be characterised
by the horizontal displacement of the tangent from the non-linear curve
(the dashed line  and the solid curve in Fig.~\ref{Fig:VvsZp})
at a given voltage.
Since the deduced change in piezo-drive position is
the same as the change in tip position in contact,
any such horizontal displacement appears directly as a change in separation.
For the measured data in Fig.~\ref{Fig:VvsZp},
at the end of the 300$\,$nm piezo-drive range in contact,
the difference between the actual piezo-drive
and the piezo-drive position at which the linear tangent at contact
gives the same  voltage is 54$\,$nm.
For the calculated data for the non-linear cantilever
used in Fig.~\ref{Fig:QvsZp-nlcant},
but over the same range of 300$\,$nm,
the difference is 1.2$\,$nm.

One can conclude from these calculations
that the non-linearity in the photo-diode
is about 50 times greater
than the non-linearity  of the cantilever
in this series of experiments.

%\newpage
%%%%%%%%%%%%%%%%%%%%%%%%%%%%%%%%%%%%%%%%%%%%%%%%%%%%%%%%%%%%
\subsection{Linear Cantilever, Non-Linear Photo-diode} \label{Sec:LinCant-nlpd}

The present series of experiments
are likely representative of the norm,
and the dominance of the photo-diode non-linearity
over the cantilever non-linearity is
likely the rule rather than the exception.
Hence it is worthwhile to give explicitly the equations required
to analyze atomic force microscope force data
in the case of a non-linear photo-diode and a linear cantilever.
These simplify the full analysis given earlier in this section.
Essentially one takes
the linear cantilever results from \S\ref{Sec:lin-cant}
and combines them with the non-linear data analysis
results of \S\ref{Sec:nl-anal}.

For the linear cantilever,
the lever arm is the constant
$L_2 = R \sqrt{ 2 +  2\cos \theta_0  } $.
Also, the deflection angle, force, and vertical tip position
are all proportional to each other,
Eqs~(\ref{Eq:QFz-lin})--(\ref{Eq:ztQ-lin}), %Eq.~(\ref{Eq:x(Q)}),
\begin{equation} \label{Eq:Q-x-z-lin}
\theta = E^\mathrm{ext} F_z ,\;
 x = D^\mathrm{ext} \theta
 \mbox{, and }
 z_\mathrm{t} = \theta /\alpha^\mathrm{ext} .
\end{equation}
The superscript `ext' ($\mu > 0$),
is replaced by `ret' ($\mu < 0$),
or by `nc' ($\mu = 0$)
in the respective regimes.

For the non-linear analysis of the measured data,
the fit to the measured raw voltage is
as in Eq.~(\ref{Eq:zp(V)}),
\begin{equation}
z_\mathrm{pc}^\mathrm{ext}(\tilde V)
=
a_0
+ a_1^\mathrm{ext} \tilde V
+ a_2^\mathrm{ext} [\tilde V ]^2
+ a_3^\mathrm{ext} [\tilde V ]^3
+ \ldots
\end{equation}
As in the linear case,
the measured voltage is fitted to a straight line
in the base-line region, Eq.~(\ref{Eq:tilde-V-b}),
\begin{equation}
\tilde V^\mathrm{ext}_\mathrm{b}(z_\mathrm{p})
=
\tilde V^\mathrm{ext}_\mathrm{b}
+ \tilde \beta^\mathrm{ext}_\mathrm{b} [z_\mathrm{p} - z_\mathrm{pb} ].
\end{equation}
Now the raw contact slope from the non-linear fit
evaluated at the base-line voltage constant,
$\tilde V^\mathrm{ext}_\mathrm{b}$
is given by Eq.~(\ref{Eq:beta-cb}),
\begin{eqnarray}
\tilde \beta_\mathrm{cb}^\mathrm{ext}
&\equiv &
\left. \frac{ \Delta \tilde V_\mathrm{c}^\mathrm{ext} }{ \Delta z_\mathrm{p} }
\right|_{\tilde V_\mathrm{b}^\mathrm{ext}}
\nonumber \\ & = &
\frac{1}{
a_1^\mathrm{ext}
+ 2 a_2^\mathrm{ext}
\tilde V^\mathrm{ext}_\mathrm{b}
+ 3 a_3^\mathrm{ext}
[\tilde V^\mathrm{ext}_\mathrm{b}]^2
+ \ldots }  .
\end{eqnarray}
The rate of change of angle with tip position
in the base-line region,
the $\alpha_\mathrm{b}^\mathrm{nc}$ of Eq.~(\ref{Eq:alpha-bnc}),
is the same as the $\alpha(0)$ of Eq.~(\ref{Eq:ztQ-lin}).
The conversion  factor for voltage to angle
in the base-line region given by Eq.~(\ref{Eq:gamma_b})
is unchanged,
$\gamma_\mathrm{b}^\mathrm{ext}
= - \tilde \beta_\mathrm{cb}^\mathrm{ext} /\alpha_\mathrm{b}^\mathrm{ext}$.

The base-line angle and the base-line angle
due to the surface force are unchanged
from Eqs~(\ref{Eq:Qb(zp)}) and (\ref{Eq:Qbf(zp)}),
\begin{equation}
\theta_\mathrm{b}^\mathrm{ext}(z_\mathrm{p})  =
E^\mathrm{nc} F^\mathrm{ext}_\mathrm{b} +
(\gamma_\mathrm{b}^\mathrm{ext})^{-1}
\tilde \beta_\mathrm{b}^\mathrm{ext}
[ z_\mathrm{p} - z_\mathrm{pb}^\mathrm{ext} ] ,
\end{equation}
and
\begin{equation}
\theta_\mathrm{bf}^\mathrm{ext}(z_\mathrm{p})  =
E^\mathrm{nc}
\left\{ F^\mathrm{ext}_\mathrm{b}
+ F^\mathrm{ext'}_\mathrm{b} [z_\mathrm{p} - z_\mathrm{pb}^\mathrm{ext}]
\right\} .
\end{equation}

The total effective angle in contact
is formally the same as Eq.~(\ref{Eq:Qtotc}),
\begin{equation}
\tilde \theta_\mathrm{tot,c}^\mathrm{ext}(\tilde V)
=
\theta_0 + \theta_\mathrm{c}^\mathrm{ext}(\tilde V)
+ \theta_\mathrm{b}^\mathrm{ext}(z_\mathrm{pc}(\tilde V))
- \theta_\mathrm{bf}^\mathrm{ext}(z_\mathrm{pc}(\tilde V)) ,
\end{equation}
with the deflection angle now being linear in the tip position,
\begin{equation}
\theta_\mathrm{c}^\mathrm{ext}(\tilde V)
=
\alpha^\mathrm{ext}
\left[ -z_0^\mathrm{ext} - z_\mathrm{pc}^\mathrm{ext}(\tilde V) \right] .
\end{equation}
At a given measured datum  $\{z_\mathrm{p},\tilde V\}$,
in or out of contact, on extension,
the deflection angle is formally the same as Eq.~(\ref{Eq:Q(zp,V)}),
\begin{equation}
\theta(z_\mathrm{p},\tilde V) =
\tilde \theta_\mathrm{tot}(\tilde V) - \theta_0
- \theta_\mathrm{b}^\mathrm{ext}(z_\mathrm{p})
+ \theta_\mathrm{bf}^\mathrm{ext}(z_\mathrm{p}) .
\end{equation}
From this,
the linear equations (\ref{Eq:Q-x-z-lin}) give the force
and also the separation, $h = z_\mathrm{p} + z_\mathrm{t} + z_0^\mathrm{ext} $.

The equations that give the constant that fixes the zero of separation
are essentially unchanged from \S \ref{Sec:zero-sep-nl}.
One has the crude estimate of the piezo-drive position
where the base-line intercepts the contact curve,
\begin{equation}
z_\mathrm{pcb}^\mathrm{ext}
=
a_0^\mathrm{ext}
+ a_1^\mathrm{ext} \tilde V^\mathrm{ext}_\mathrm{b}
+ a_2^\mathrm{ext}
[\tilde V^\mathrm{ext}_\mathrm{b}]^2
%\nonumber \\ && \mbox{ }
+ a_3^\mathrm{ext}
[\tilde V^\mathrm{ext}_\mathrm{b} ]^3
+ \ldots ,
\end{equation}
and the refined version, Eq.~(\ref{Eq:zpcb*}),
\begin{equation}
z_\mathrm{pcb}^\mathrm{ext,*}
=
\frac{\tilde \beta_\mathrm{cb}^\mathrm{ext} z_\mathrm{pcb}^\mathrm{ext}
- \tilde \beta_\mathrm{b}^\mathrm{ext} z_\mathrm{pb}^\mathrm{ext}
}{\tilde \beta_\mathrm{cb}^\mathrm{ext}
- \tilde \beta_\mathrm{b}^\mathrm{ext} } .
\end{equation}
Equation (\ref{Eq:qf(zpb)}) again follows,
$ \theta_\mathrm{c}^\mathrm{ext}(\tilde V^*)
=
\theta_\mathrm{bf}^\mathrm{ext}(z_\mathrm{pcb}^\mathrm{ext,*}) $,
which in the present case can be rearranged as
\begin{equation}
z_0^\mathrm{ext}
=
-z_\mathrm{pcb}^\mathrm{ext,*}
-
[ E^\mathrm{nc} /\alpha^\mathrm{ext} ]
\left\{ F^\mathrm{ext}_\mathrm{b}
+ F^\mathrm{ext'}_\mathrm{b}
[z_\mathrm{pcb}^\mathrm{ext,*} - z_\mathrm{pb}^\mathrm{ext}]
\right\} .
\end{equation}
This exactly the same as given in \S \ref{Sec:zero-sep-nl},
as one might expect since linear analysis suffices for the base-line.

%\newpage
%%%%%%%%%%%%%%%%%%%%%%%%%%%%%%%%%%%%%%%%%%%%%%%%%%%%%%%%%%%%%%%%%%%%
%
\section{Results} \label{Sec:results}
%
%%%%%%%%%%%%%%%%%%%%%%%%%%%%%%%%%%%%%%%%%%%%%%%%%%%%%%%%%%%%%%%%%%%%

%%%%%%%%%%%%%%%%%%%%%%%%%%%%%%%%%%%%%%%%%%%%%%%%%%%%%%%%%%%%%%%%%%
\begin{figure}[t!]
\centerline{
\resizebox{8.5cm}{!}{ \includegraphics*{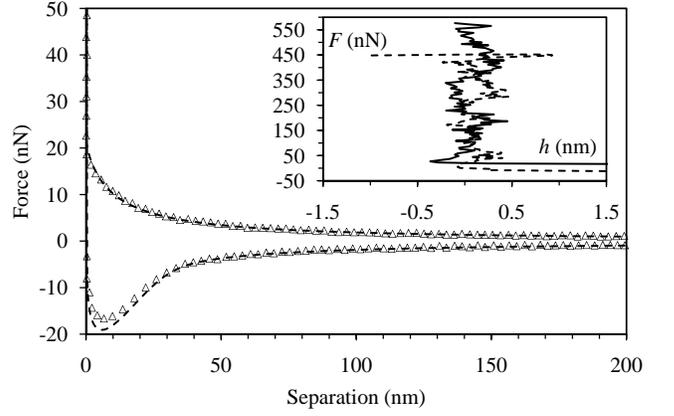} } }
% From My Documents\Papers\Current\nonlinear AFM\nl-SiSi2.xlsx:Fig4
\caption{\label{Fig:Fvsh2}
Measured\cite{Zhu12,expt} and calculated force versus separation
for a drive velocity of $\dot z_\mathrm{p} = 2\,\mu$m$\,$s$^{-1}$,
viscosity $\eta = 51.62\,$mPa$\,$s$^{-1}$,
friction coefficient $\mu = 0.35$,
and cantilever spring constant $k_0 = 1.37\,$N/m.
The symbols are the atomic force microscope data
(upper is for extension, lower is for retraction),
analyzed with the non-linear photo-diode, linear cantilever algorithm.
The almost overlapping dashed curves are the calculated drainage force
with a slip length of $b=3\,$nm
and effective spring constant of $k_\mathrm{eff} = 1.68\,$N/m.
The right inset magnifies the measured data in contact
(jagged solid curve is extension, jagged dashed curve is retraction).
}
\end{figure}
%%%%%%%%%%%%%%%%%%%%%%%%%%%%%%%%%%%%%%%%%%%%%%%%%%%%%%%%%%%%%%%%%%

%%%%%%%%%%%%%%%%%%%%%%%%%%%%%%%%%%%%%%%%%%%%%%%%%%%%%%%%%%%%%%%%%%
\begin{figure}[t!]
\centerline{
\resizebox{8.5cm}{!}{ \includegraphics*{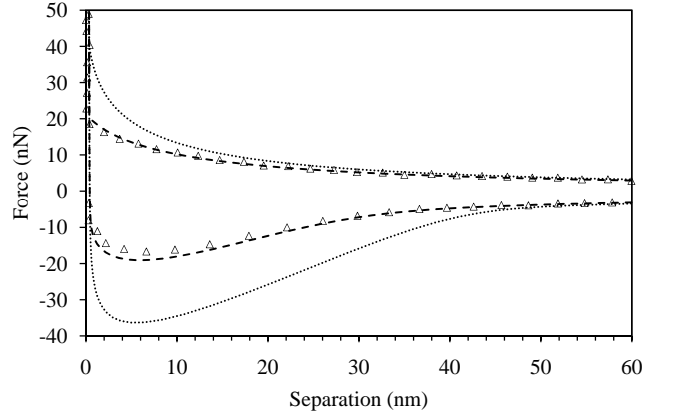} } }
% From My Documents\Papers\Current\nonlinear AFM\nl-SiSi2.xlsx:Fig5
\caption{\label{Fig:Fvsh2a}
Magnification of previous figure at small separations.
The symbols and dashed curves are as in the preceding figure,
and the dotted curves are the calculated stick drainage force
($b=0\,$nm, $k_\mathrm{eff} = 1.68\,$N/m).
}
\end{figure}
%%%%%%%%%%%%%%%%%%%%%%%%%%%%%%%%%%%%%%%%%%%%%%%%%%%%%%%%%%%%%%%%%%

Figure \ref{Fig:Fvsh2} shows the force versus separation data
that corresponds to the raw measured data presented in Fig.~\ref{Fig:VvsZp}.
These and all of the experimental results presented in this section
are analysed using the non-linear photo-diode, linear cantilever algorithm
presented in the preceding subsection,
unless explicitly stated otherwise.
Four coefficients were used in each of the non-linear fits,
$a_0$, $a_1$, $a_2$, and $a_3$ on each branch.
The measurements in Fig.~\ref{Fig:Fvsh2} were performed
at low velocity, $\dot z_\mathrm{p} = 2\,\mu$m$\,$s$^{-1}$,
and so the magnitude of the drainage force is comparatively weak.

The figure includes the calculated drainage force,
which is given by the Taylor equation with slip,
\begin{equation}
F = \frac{-6\pi\eta R^2 \dot h}{h} f(h) ,
\end{equation}
where the slip factor is
\cite{Vinogradova95}
\begin{equation}
f(h) = \frac{h}{3b} \left[
\left( 1 + \frac{h}{6b} \right)
\ln \left( 1 + \frac{6b}{h} \right) - 1 \right] ,
\end{equation}
with $b$ being the slip length.
The case $b=0$ (equivalently, $f(h) = 1$)
corresponds to stick boundary conditions.

A weak van der Waals force
(Hamaker constant $1.3\times 10^{-21}\,$J$\,$m$^{-2}$)
and short range repulsion (equilibrium separation 0.53$\,$nm)
derived from a Lennard-Jones 6-12 potential
was included in order to calculate results in contact.
These have no effect for separations greater than about 1$\,$nm.
Details of the algorithm used to calculate the drainage force
may be found in Ref.~\onlinecite{Zhu11a}.

Comprehensive atomic force microscope
measurements of the drainage force and of the slip length
have been given in earlier work by the present authors,
\cite{Zhu12,Zhu11a,Zhu11b}
albeit with the linear analysis.
The focus in the present work is not directly on the drainage force
or the slip length,
but rather on those aspects of the measurements
where the non-linear analysis makes a difference.

Accordingly, one can briefly observe
that the agreement between theory and measurement
in Fig.~\ref{Fig:Fvsh2} indicates the overall validity of the non-linear
procedures developed here.
The cantilever  spring constant, $k_0 = 1.37\,$N/m,
was obtained by fitting by eye the data and the calculated force
at larger separations where the stick and slip theories overlap.
In fact, at this velocity, $1.3 \alt k_0 \alt 1.6 \,$N/m
gave equally good fits.
At the higher velocity $\dot z_\mathrm{p} = 20\,\mu$m$\,$s$^{-1}$,
$1.3 \alt k_0 \alt 1.5\,$N/m were acceptable,
and at the still higher velocity  $\dot z_\mathrm{p} = 50\,\mu$m$\,$s$^{-1}$,
$1.33 \alt k_0 \alt 1.42\,$N/m fitted the data.
(The higher velocities gave larger drainage forces
and the fits were more definitive.
These cases are discussed in detail below.)
Based on these fits,
the value of the cantilever  spring constant
is taken to be $k_0 = 1.37\,$N/m, here and below.
It is estimated that the value has been obtained
with a precision of about $\pm 0.1\,$N/m.
The effective spring constant corresponding to this
is given by Eq.~(\ref{Eq:keff}), $k_\mathrm{eff} = 1.68\,$N/m.
This effective spring constant should be used
in the conventional linear analysis of the data,
and in the theoretical calculations that
use a simple spring model.

%%%%%%%%%%%%%%%%%%%%%%%%%%%%%%%%%%%%%%%%%%%%%%%%%%%%%%%%%%%%%%%%%%
\begin{figure}[t!]
\centerline{
\resizebox{8.5cm}{!}{ \includegraphics*{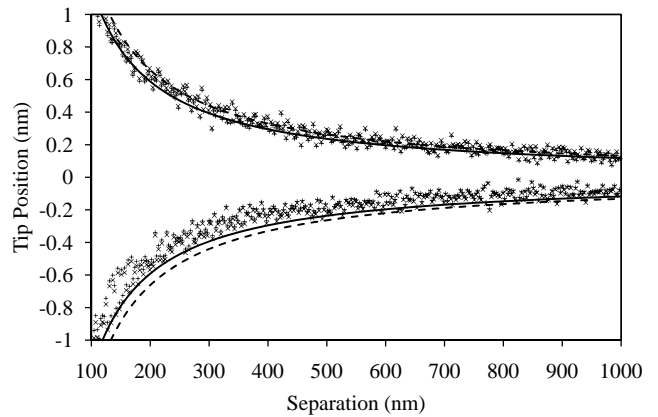} } }
% From My Documents\Papers\Current\nonlinear AFM\nl-SiSi2.xlsx:zvsh
\caption{\label{Fig:zvsh2}
Fit for the spring constant,
(data corresponding to Fig.~\ref{Fig:Fvsh2}).
The vertical tip position, $z_\mathrm{t}=F/k_\mathrm{eff}$, is shown.
The plus symbols are the non-linear analysis
of the raw experimental data,\cite{Zhu12,expt}
the cross symbols are the original linear analysis
of the same data,\cite{Zhu12}
the solid curve is the position calculated
from the stick  drainage force
($b=0\,$nm, $k_\mathrm{eff} = 1.68\,$N/m),
and the dashed curve is the position calculated
from the stick  drainage force
($b=0\,$nm, $k_\mathrm{eff} = 1.5\,$N/m).
}
\end{figure}
%%%%%%%%%%%%%%%%%%%%%%%%%%%%%%%%%%%%%%%%%%%%%%%%%%%%%%%%%%%%%%%%%%

Figure \ref{Fig:zvsh2} shows the quality of the fit
used to obtain the cantilever spring constant
at the lowest drive velocity.
It can be seen that a value of
$k_0 = 1.37\,$N/m (equivalently, $k_\mathrm{eff} = 1.68\,$N/m)
slightly overestimates the magnitude of the deflection,
more noticeably on the retraction branch,
and that a slightly higher value would give a better fit in this case.
Note that in order to reduce the fitting parameters from two to one,
only the stick theory is used for the fit.
(In any case the slip theory is virtually coincident
at these large separations.)
Also included is the original linear analysis
of the same raw data,\cite{Zhu12}
and the calculated stick drainage force
that was used to fit a value of $k_\mathrm{eff} = 1.5\,$N/m
(equivalently, $k_0 = 1.23\,$N/m).
First it may be observed that there is only a minor difference
between the linearly and the non-linearly analyzed experimental data;
the linear data slightly overestimates the magnitude
of the deflection, and this effect increases
as the magnitude of the force increases.
Second, the value of the spring constant used originally
is about 10\% smaller than that fitted here.
This gives a slightly better fit to the linearly analyzed extend data,
but a worse fit on the retract curve, which was not taken into account
in Ref.~\onlinecite{Zhu12}.
At smaller separations than shown,
the original and the present fits slightly improve,
but in this region the slip length begins to have an effect
and so it is better not to include it in the determination
of the spring constant.
Although the discrepancy in the spring constant is small
(errors in other methods of cantilever calibration typically exceed 20\%),
it does lead to a large discrepancy in the fitted slip length
and to qualitatively different behavior of the slip length
with shear rate to that found here.

The overlap between the measured data and the slip calculations
at short range in Figs~\ref{Fig:Fvsh2} and  \ref{Fig:Fvsh2a}
indicates that the correct slip length,
$b=3\,$nm, has been obtained.
The agreement is surprisingly good for the retract data,
which was not analyzed in Refs~\onlinecite{Zhu12,Zhu11a,Zhu11b},
but which turns out to be very sensitive to the slip length.
For example, the adhesive force,
which is defined to be the minimum
of the retract force,
in Fig.~\ref{Fig:Fvsh2a} is measured to be $-16.6\,$nN,
and it is calculated to be $-19.1\,$nN  for $b=3\,$nm
For the case of stick ($b=0\,$nm),
it is calculated to be $-36.3\,$nN,
which significantly overestimates the adhesion.
The extend data is less sensitive to the slip length.
For example, at $h=10\,$nm,
the measured force is 10.5$\,$nN,
the calculated slip force for  $b=3\,$nm is 10.1$\,$nN,
and the calculated stick force, $b=0\,$nm, is 13.2$\,$nN.
Based on this and the higher velocity results shown below,
it is estimated that the slip length has been  obtained
with a precision of about $\pm 0.5\,$nm.

The linear analysis (not shown) of the measured extend data
using the slope of the tangent at first contact
gives 10.8$\,$nN at $h=10\,$nm,
and, using the slope of the tangent at final contact
it gives $-16.5\,$nN for the adhesion.
Hence the error due to the linear analysis
is quite small at these points at this lowest velocity.
It should be noted that the linear results
were obtained using the effective spring constant
$k_\mathrm{eff} = 1.68\,$N/m
rather than the cantilever spring constant,
$k_0 = 1.37\,$N/m.

It is worth mentioning that the drainage measurements
reported in Ref.~\onlinecite{Zhu12}
found a low shear rate limiting slip length
of $b_0 = 10\,$nm for the Si-Si case.
This is significantly larger than is found here.
In that investigation it was found that
the slip theory progressively underestimated the measured force
at small separations and overall it was a noticeably worse fit
than found here.
As mentioned above,
the retract data was not included in the fit in Ref.~\onlinecite{Zhu12}.
The most likely reason for the larger slip length
and the degradation in the fit at small separations in
Ref.~\onlinecite{Zhu12} compared to here
is the difference in the value of the spring constant used:
here $k_\mathrm{eff} = 1.68\,$N/m,
compared to  $k_\mathrm{eff} = 1.5\,$N/m in Ref.~\onlinecite{Zhu12}.
A detailed discussion of how the linear analysis
led to the contradiction between these earlier results
and the present ones is given in the concluding section.

%As mentioned above,
%the spring constant was fitted in Ref.~\onlinecite{Zhu12}
%using the linear data analysis,
%and this is part of the reason for underestimating it.
%Other possibilities
%(neglect of friction, linear data analysis, ambiguity in zero of separation,
%ambiguity in photo-diode calibration factor)
%do not appear to have as big an effect as the value of the spring constant.

The friction coefficient was determined
from the measured data in Fig.~\ref{Fig:Fvsh2}
by equalizing the extension and retraction calibration factors
using the
Guided Unbiased Estimate for a Single Solution algorithm,
which is a rapid and well-used procedure for solving non-linear equations.
A value of $\mu = 0.38$ gave
$\gamma^\mathrm{ext}_\mathrm{cb} = 6357\,$V/rad
and
$\gamma^\mathrm{ret}_\mathrm{cb} = 6359\,$V/rad.
This is a little larger than
the value determined by Stiernstedt et al.,\cite{Stiernstedt05}
who obtained $\mu = $ 0.32--0.35 in four independent friction measurements
(two probes, axial and lateral methods)
for a silica probe on a silica substrate.
In view of this a value of $\mu =0.35$ was used to analyse
the data in Figs \ref{Fig:Fvsh2} and \ref{Fig:Fvsh2a}
and all of the following figures.

It was found that the value of the friction coefficient
varied somewhat with the voltage at which $\gamma$
was evaluated.
Obtaining $\mu$ by minimising
$[\gamma^\mathrm{ext}(V;\mu)-\gamma^\mathrm{ret}(V;\mu)]^2$
over the contact region also gave a different value.
In some case, particularly at higher drive velocities,
unphysical values of $\mu$ were obtained.
It was concluded that for these particular measurements
in this high viscosity liquid
the non-linear method could not be used
to obtain the friction coefficient reliably.
As is discussed further below,
the most likely reason for the failure
of this method of measuring the friction coefficient
in the present case is the neglect of the variations in the  drag force,
which are most pronounced in contact.

\comment{ %%%%%%%%%%%%%%%%%%%%%%%%%%%%%%%%%%%%%%%%%%%%%%%%%%%%%%%
As a comparison,
a value of $\mu = 0.40$ gave
$\gamma^\mathrm{ext} = 6365\,$V/rad
and
$\gamma^\mathrm{ret} = 6354\,$V/rad,
and a value of $\mu = 0.30$ gave
$\gamma^\mathrm{ext} = 6324\,$V/rad
and
$\gamma^\mathrm{ret} = 6379\,$V/rad.

The friction coefficient was found to be relatively sensitive
to the choice of contact region for the non-linear fit,
particularly at the low voltage end.
The above result of $\mu = 0.38$ was obtained by a fit on the range
$z_\mathrm{p}^\mathrm{ext} \in [1485,1213]$ (in nm).
Using instead a range $z_\mathrm{p} \in [1499,1213]$
gave $\mu = 0.46$,
and a range $z_\mathrm{p}^\mathrm{ext} \in [1470,1213]$
gave $\mu = 0.45$.
(First contact, which is the point of intersection
of the extrapolation of the contact and the base-line fits,
was $z^\mathrm{ext,*}_\mathrm{pcb} = 1507\,$nm.)
One can say that the estimate of the friction coefficient
has an error of about $\pm 0.1$.
} % ed comment %%%%%%%%%%%%%%%%%%%%%%%%%%%%%%%%%%%%%%%%%%

Interestingly enough,
the choice of the friction coefficient
(and the choice of the fitting region)
had almost no effect on the values of the measured
non-contact forces.
This is discussed further below,
but in essence,
the method of analyzing the non-contact data
cancels the friction contribution whatever it may be.

The inset of Fig.~\ref{Fig:Fvsh2}
shows the force versus separation just prior to and in contact.
Overall in contact the non-linear analysis gives a quite vertical curve
that fluctuates about $h=0$ with apparent noise
on the order of $\pm 0.5\,$nm.
The verticality of the contact region
of the data analysed with the non-linear algorithm
is much better than that obtained with the linear analysis.
As can be seen in the inset to  Fig.~\ref{Fig:VvsZp},
the linear analysis gives
a systematic error in contact of 20--50$\,$nm.

This noise in the non-linear data
is not entirely random or unphysical
(hence the word `apparent')
because any roughness of the surface
gives rise to physical fluctuations in the apparent separation,
as is discussed quantitatively shortly.
In fact the presentation of the data in Fig.~\ref{Fig:Fvsh2}
exaggerates the apparent noise in the non-linear analysis.
This is because the friction force is reversed
on the retract branch compared to the extend branch,
and so a given surface force $F$ corresponds to different
contact positions on the substrate.
Hence the disagreement between the extend and retract traces
when presented in the form of force versus separation
is entirely as one would expect.

%%%%%%%%%%%%%%%%%%%%%%%%%%%%%%%%%%%%%%%%%%%%%%%%%%%%%%%%%%%%%%%%%%
\begin{figure}[t!]
\centerline{
\resizebox{8.5cm}{!}{ \includegraphics*{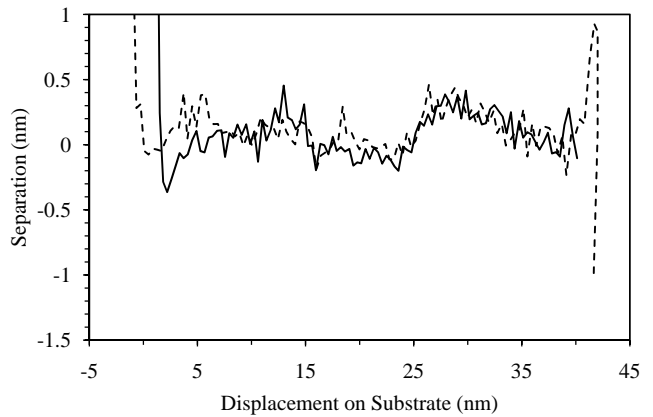} } }
% From My Documents\Papers\Current\nonlinear AFM\nl-SiSi2.xlsx:Fig6
\caption{\label{Fig:hvsy2}
Measured\cite{Zhu12,expt} separation versus
displacement on substrate in contact
corresponding to Fig.~\ref{Fig:Fvsh2}.
The solid curve is extension and the dashed curve is retraction.
}
\end{figure}
%%%%%%%%%%%%%%%%%%%%%%%%%%%%%%%%%%%%%%%%%%%%%%%%%%%%%%%%%%%%%%%%%%

One needs to compare the two branches at the same physical position
on the substrate.
The basis of the analysis of the tilted cantilever
is that in contact the probe slides along the substrate
in the axial direction of the cantilever,
which is the origin of the friction force.
The displacement from first contact of the probe
on the substrate is simply $y = L_2 \theta$.
(Actually, the full expression is
$y = S_0 x - C_0 L_2 \theta
= [ S_0 D^\mathrm{ext/ret} - C_0 L_2 ]\theta$,
but this is not used here.)
%(Actually, this is the displacement from the point at
%which the angle would be zero, which is close to the point at which
%the extrapolated contact and  base-line  raw voltages intersect.)
This is used to plot the apparent separation,
$h = z_\mathrm{p} + z_\mathrm{t} + z_0$,
for extend and retract in Fig.~\ref{Fig:hvsy2}.
As in the preceding figure,
the linear cantilever, non-linear photo-diode analysis was used
to obtain $z_\mathrm{t}(\theta(\tilde V, z_\mathrm{p}))$.

As mentioned in the text
(c.f.\ the fourth paragraph of \S\ref{Sec:z0-lin}),
the separation in contact
measures the difference between changes in the piezo-drive position
and changes in the tip position.
It is positive when there is a protuberance on the substrate,
and it is negative when there is a depression.
That this is really a topographic map of the substrate
can be seen by the high degree of correlation
between the extend and the retract traces in Fig.~\ref{Fig:hvsy2}.
There are slight depressions, about 0.2$\,$nm deep,
at about $y=20\,$nm and $y=38\,$nm,
and there is a protuberance, about 0.5$\,$nm high, at about $y=30\,$nm.
Where there is no correlation between the extend and retract traces,
one can attribute the departure from zero to noise
that is most likely connected with stick-slip motion
of the probe on the substrate.
One can conclude that the substrate is really quite smooth and,
despite this noise,
the method does give reliable topographic information
with a resolution on the order of 0.1$\,$nm.
%It is to be noted that the linear analysis gives an error
%of about 50$\,$nm for the separation in contact
%(c.f.\ the inset of  Fig.~\ref{Fig:VvsZp}).

It was found that $z_0^\mathrm{ext} = -1506.9\,$nm
and that  $z_0^\mathrm{ret} = -1507.8\,$nm.
These differ by 1$\,$nm,
probably due to the reversal of the friction force on the two branches,
which causes a displacement of the vertical position of the tip.
This 1$\,$nm difference appears to be essential to get the very precise
overlap on the topographic plot.
It is emphasized that the two values of $z_0$ were generated by
the algorithm and no human judgement or adjustment was involved.

The rather pronounced feature
in the start of the retract curve at about $y=42\,$nm in Fig.~\ref{Fig:hvsy2}
is not a physical feature.
It is an artefact of the reversal of direction
of the piezo-drive on the change from extend to retract.
This reversal begins at the start of the retract branch
in this particular model of the atomic force microscope.
At this turning point,
the probe is essentially stuck at one contact position
as the friction force reverses direction.
During this turning phase the present analysis,
which assumes that $F_y = - \mu F_z$,
is not valid.
For the same reason the negative separation of about $h\approx -0.5\,$nm
just after first contact on extension is likely an artifact
of the linear friction model.

%%%%%%%%%%%%%%%%%%%%%%%%%%%%%%%%%%%%%%%%%%%%%%%%%%%%%%%%%%%%%%%%%%
\begin{figure}[t!]
\centerline{
\resizebox{8.5cm}{!}{ \includegraphics*{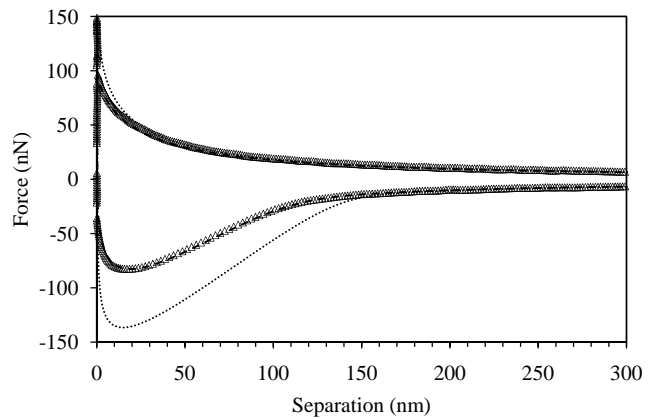} } }
% From My Documents\Papers\Current\nonlinear AFM\nl-SiSi20.xlsx:Fig7
\caption{\label{Fig:Fvsh20}
Measured\cite{Zhu12,expt} and calculated force versus separation
for a drive velocity of $\dot z_\mathrm{p} = 20\,\mu$m$\,$s$^{-1}$
and viscosity $\eta = 52.25\,$mPa$\,$s$^{-1}$.
The symbols are the atomic force microscope data
(upper is for extension, lower is for retraction),
analyzed with the non-linear photo-diode, linear cantilever algorithm
with cantilever spring constant $k_0=1.37\,$N/m
and with friction coefficient of $\mu = 0.35$.
The almost completely obscured dashed curves are the calculated drainage force
with a slip length of $b = 3\,$nm and
$k_\mathrm{eff} =1.68\,$N/m.
The dotted curves are the calculated drainage force
for stick boundary conditions
($b = 0\,$nm and $k_\mathrm{eff} =1.68\,$N/m).
}
\end{figure}
%%%%%%%%%%%%%%%%%%%%%%%%%%%%%%%%%%%%%%%%%%%%%%%%%%%%%%%%%%%%%%%%%%

Figure \ref{Fig:Fvsh20} shows the measured and calculated drainage force
for a drive velocity of $\dot z_\mathrm{p} = 20\,\mu\,$m$\,$s$^{-1}$,
which is ten times the velocity of the preceding case.
The viscosity $\eta = 52.25\,$mPa$\,$s$^{-1}$
has slightly increased due to a temperature rise in the measurement cell.
The cantilever spring constant was unchanged, $k_0=1.37\,$N/m.
The calculated slip force (obscured dashed curves)
was obtained with a simple spring model and
the unchanged effective spring constant of
$k_\mathrm{eff} =1.68\,$N/m and unchanged slip length $b=3\,$nm.
The fact that an unchanged spring constant
and slip length fit the measured data equally well
at this high velocity as at the preceding low velocity
indicates the reality of the values used,
the validity of the mathematical form for the drainage force,
and the reliability of the non-linear algorithm
for the conversion of the raw measured signal into quantitative data.

The fact that the slip length is the same
here as for the low velocity case
(and for the higher velocity of
$\dot z_\mathrm{p} =  50\,\mu\,$m$\,$s$^{-1}$
presented below)
suggests that the slip length is independent of the shear rate.
This conclusion is reinforced by the fact
that the slip theory with constant slip length
fits the measured data over all separations
(c.f.\ Figs  \ref{Fig:Fvsh2a} and \ref{Fig:Fvsh20}),
since the maximum shear rate of the drainage flow
increases with decreasing separation.
This contrasts with earlier work
in which it was found that the slip theory
increasingly underestimated the measured drainage force
on extension as the separation approached zero.\cite{Zhu12a}
As was discussed above,
%in the present series of measurement
such behavior occurs when  the spring constant is underestimated,
which was the case in Ref.~\onlinecite{Zhu12}
from which  series of measurements the raw data used here was taken.
It is not known at this time whether or not a similar underestimate
occurred in Ref.~\onlinecite{Zhu12a}.
Accordingly, the claim that the  slip length decreases with increasing
shear rate\cite{Zhu12a}
should be treated with caution
pending  a more refined analysis of that data.

The adhesion measured using the non-linear analysis is $-82.6\,$nN,
and that calculated with the slip length $b=3\,$nm is $-82.6\,$nN.
By way of comparison,
the adhesion measured using the linear analysis
and $k_\mathrm{eff} =1.68\,$N/m is $-80.8\,$nN (not shown),
and that calculated with the stick theory  $b=0\,$nm is $-136.6\,$nN.
Note that it is essential to use the effective spring constant
in order to get such reliable linear results.
%The error in the adhesion that arises from the linear analysis
%(with the effective spring constant) is about 3\%.

\begin{figure}[t!]
\centerline{
\resizebox{8.5cm}{!}{ \includegraphics*{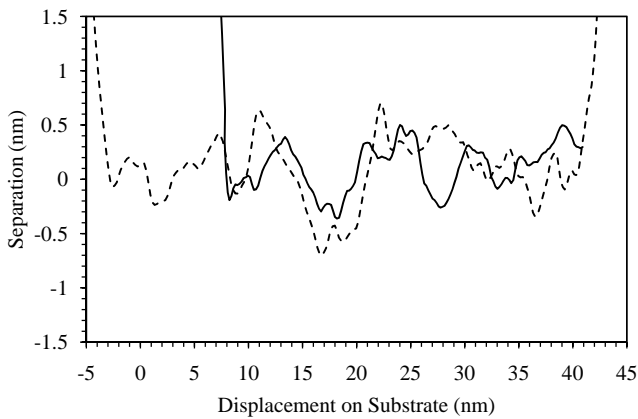} } }
% From My Documents\Papers\Current\nonlinear AFM\nl-SiSi20.xlsx:Fig8
\caption{\label{Fig:hvsy20}
Measured\cite{Zhu12,expt} separation versus
displacement on substrate in contact,
(corresponding to preceding figure).
The solid curve is extension and the dashed curve is retraction.
}
\end{figure}

The apparent separation  in contact as a function of position
on the substrate is shown in Fig.~\ref{Fig:hvsy20}.
There is a degree of correlation between the extend and retract traces,
with a noticeable crater appearing in both at about $y=20\,$nm.
The amount of noise is noticeably larger
at this velocity of $\dot z_\mathrm{p} = 20\,\mu$m$\,$s$^{-1}$
compared to the data in Fig.~\ref{Fig:hvsy2},
which was obtained at a velocity of $\dot z_\mathrm{p} = 2\,\mu$m$\,$s$^{-1}$.
The physical origin of the increased noise is likely that
more power is dissipated at the greater velocity.
Two signatures of this change can be noticed.
First is the reduced correlation between the extend and the retract traces.
And second is that the amplitude of the fluctuations
appear larger.
The spatial wave length of the fluctuations is noticeably larger
in Fig.~\ref{Fig:hvsy20} compared to  Fig.~\ref{Fig:hvsy2}.
This indicates that the fluctuations are dynamic in nature,
and that they probably have the same temporal frequency in both cases.
This point explains why the low velocity data
is more satisfactory in producing a topographic map of the surface:
at low velocities
the high frequency vibrations of the cantilever
are averaged out in the time it takes to traverse a given spatial feature,
whereas at high velocities the period of vibration
and the spatial period are comparable and so they interfere with each other.
One can conclude that the low velocity data gives more reliable results
for the contact region than the high velocity case.

%%%%%%%%%%%%%%%%%%%%%%%%%%%%%%%%%%%%%%%%%%%%%%%%%%%%%%%%%%%%%%%%%%
%%%%%%%%%%%%%%%%%%%%%%%%%%%%%%%%%%%%%%%%%%%%%%%%%%%%%%%%%%%%%%%%%%

This last point is important.
The friction coefficient used in the non-linear analysis
in the high velocity case was
$\mu = 0.35$, which was taken from the low velocity data.
(A non-linear fit in the contact region was still performed and used
for the high velocity case.)
This value of $\mu$
in the present case of $\dot z_\mathrm{p} = 20\,\mu$m$\,$s$^{-1}$
gives calibration coefficients
$\gamma^\mathrm{ext}_\mathrm{cb}  = 5660\,$V/rad
and  $\gamma^\mathrm{ret}_\mathrm{cb} = 6425\,$V/rad.
(In the analysis of the data,
these  values of $\gamma$ were used on each respective branch.)
An unrealistically high value of $\mu = 1.3$ is required
to bring these into agreement at $\gamma_\mathrm{cb} = 6250\,$V/rad.
Although the required friction coefficient is unrealistically high,
using it only changed the calibration coefficient by about 3\%;
the measured value of the adhesion remained unchanged at  $-82.6\,$nN.

Besides the increased noise due to cantilever vibration,
there are two further reasons why the high velocity data
is unreliable in contact.
First, in addition to the friction force,
an axial drag force acts on the probe parallel to the substrate
as it slides on the substrate.
This axial drag force is not taken into account
in the present analysis.
It is exacerbated in high viscosity liquids and at high velocities.
Second, the  drag force normal to the cantilever,
which is distinct from the drainage force,
is not in fact constant as assumed here,
but varies with load.\cite{Zhu11a,Zhu11b}
(For determining the slip length,
it is not necessary to use here a variable drag model
because  the cantilever is relatively stiff.)\cite{Zhu11a,Zhu11b}
This variable drag effect increases with increasing viscosity,
increasing velocity, and increasing force.
Although for stiff cantilevers such as the present
the variation in drag is relatively negligible in the non-contact region,
in contact the variation is independent of the stiffness of the cantilever
and so it can be expected to affect the present results.
A simple calculation shows that here the axial drag force
is several orders of magnitude smaller than the friction force.
Hence it is most likely that
it is the present neglect of the variable drag force
in contact that makes the determination of the friction coefficient
in the present series of measurements unreliable.

The question naturally arises
that if the friction coefficient is unreliable for a given velocity,
why can one use the non-linear contact fits at that velocity?
There are two answers to this, one pragmatic and one reasoned.
First, the evidence is that the fits are reliable and produce
quantitative measured results,
namely that the measured adhesion is insensitive to the value of
the friction coefficient used in the analysis,
and the measured force agrees quantitatively with the calculated force.
Second,
the non-linearity in the raw voltage measured in contact
arises almost entirely from the photo-diode;
the cantilever, including the friction, contributes to the linear part.
The two drag forces just mentioned and neglected here also contribute to
the linear part.
Similar to the friction force, they reverse sign between extend and retract.
The method of analyzing the non-contact data
essentially cancels the friction contribution,
and so it would also have canceled
these drag contributions if they had been taken into account.
This is the justification for using the friction coefficient
measured at low velocities in combination with the non-linear fits
applied at the current velocity.

%%%%%%%%%%%%%%%%%%%%%%%%%%%%%%%%%%%%%%%%%%%%%%%%%%%%%%%%%%%%%%%%%%
\begin{figure}[t!]
\centerline{
\resizebox{8.5cm}{!}{ \includegraphics*{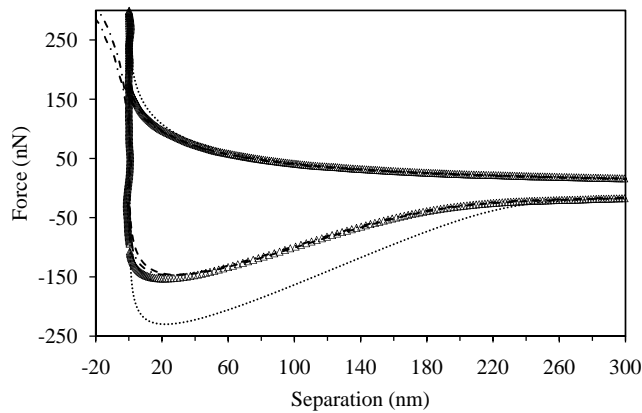} } }
% From My Documents\Papers\Current\nonlinear AFM\nl-SiSi50.xlsx:Fig9
\caption{\label{Fig:Fvsh50}
Measured\cite{Zhu12,expt} and calculated force versus separation
for a drive velocity of $\dot z_\mathrm{p} = 50\,\mu$m$\,$s$^{-1}$
and viscosity $\eta = 52.25\,$mPa$\,$s$^{-1}$.
The symbols are the atomic force microscope data
(upper is for extension, lower is for retraction),
analyzed with the non-linear photo-diode, linear cantilever algorithm
with cantilever spring constant $k_0=1.37\,$N/m
and with friction coefficient of $\mu = 0.35$.
The mainly obscured dashed curves are the calculated drainage force
with a slip length of $b = 3\,$nm and
$k_\mathrm{eff} =1.68\,$N/m.
The dotted curves are the calculated drainage force
for stick boundary conditions
($b = 0\,$nm and $k_\mathrm{eff} =1.68\,$N/m).
The obscured dash-dotted curves are measured data analyzed with the conventional
linear approach using $k_\mathrm{eff} =1.68\,$N/m,
with the zero of separation established as described in \S\ref{Sec:z0-lin}.
}
\end{figure}
%%%%%%%%%%%%%%%%%%%%%%%%%%%%%%%%%%%%%%%%%%%%%%%%%%%%%%%%%%%%%%%%%%

Figure \ref{Fig:Fvsh50}
shows the measured and calculated force
for a drive velocity of $\dot z_\mathrm{p} = 50\,\mu$m$\,$s$^{-1}$.
The measured data are analyzed
with the non-linear algorithm of \S\ref{Sec:LinCant-nlpd}
(symbols),
and with the linear algorithm  of \S\ref{Sec:Lin-tilt}
using the tangent at first (extension)
or last (retraction) contact (dash-dotted curves).
The cantilever spring constant, $k_0=1.37\,$N/m,
slip length, $b=3\,$nm,
and friction coefficient, $\mu=0.35$,
have been fixed at the values determined
in the low velocity case, $\dot z_\mathrm{p} = 2\,\mu$m$\,$s$^{-1}$.

With this value of $\mu$,
in the present case of $\dot z_\mathrm{p} = 50\,\mu$m$\,$s$^{-1}$
the calibration coefficients are
$\gamma^\mathrm{ext}_\mathrm{cb} = 6263\,$V/rad
and  $\gamma^\mathrm{ret}_\mathrm{cb} = 6416\,$V/rad.
A value of $\mu=0.57$ gives
$\gamma^\mathrm{ext}_\mathrm{cb} = 6361\,$V/rad
and  $\gamma^\mathrm{ret}_\mathrm{cb} = 6365\,$V/rad.

The adhesion measured using the non-linear analysis is $-153\,$nN,
that measured using the linear analysis is $-147\,$nN ,
that calculated with the slip length $b=3\,$nm is $-147\,$nN,
and that calculated with the stick theory  $b=0\,$nm is $-230\,$nN.

In the case of the linear analysis,
the zero of separation was established as described in \S\ref{Sec:z0-lin}.
This removes any human intervention or bias,
which can be a problem in the conventional linear analysis
where each curve is shifted horizontally
so that it gives $h=0$ at what appears by eye to be first or last contact.
In fact, in the absence of the non-linear result
it would have been difficult to establish contact with any confidence
in this case due to the smooth and continuous nature of the forces
and the lack of verticality in contact in the linearly analyzed data.
These results with negative gradient
and negative separation for the linear analysis in contact
is a significant failing of the approach.

The verticality of the linear analysis in contact
for forces $F_z \in [-150\,$nN,$150\,$nN$]$,
suggests that the photo-diode is linear in this range.
Since the non-contact forces also lie in this range,
one can understand why the linear analysis
with effective spring constant works in this high velocity case.

It can be noted in Fig.~\ref{Fig:Fvsh50}
that the maximum magnitude of the drainage forces out of contact
are about on the limit of photo-diode linearity.
At higher velocities they will enter the non-linear regime,
and so one would expect the linear analysis to become progressively
less reliable as the velocity is increased.

It can also be seen that on the extension branch,
in this figure and in the preceding figures,
there is no overlap between the contact voltages
and the non-contact voltages.
However, the retraction branch in contact encompasses
all of the non-contact voltages.
Whereas extrapolation of the contact fit is required on extension,
only interpolation is required on retraction branch.

There is a noticeable discontinuity
in the non-linear measured data at final contact on retraction,
and at first contact on extension (obscured).
This is an artefact of the linear friction model,
which is zero out of contact, and jumps discontinuously to a non-zero
value in contact. Whenever the load is non-zero at such a point,
the friction force is discontinuous and it gives rise
to a discontinuity in the analyzed data.

%%%%%%%%%%%%%%%%%%%%%%%%%%%%%%%%%%%%%%%%%%%%%%%%%%%%%%%%%%%%%%%%%%
\begin{figure}[t!]
\centerline{
\resizebox{8.5cm}{!}{ \includegraphics*{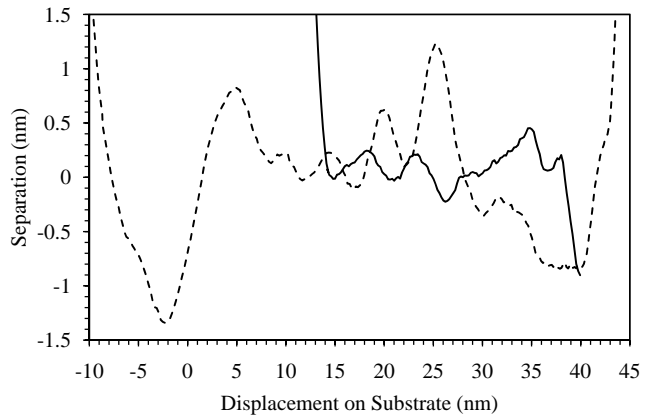} } }
% From My Documents\Papers\Current\nonlinear AFM\nl-SiSi50.xlsx:Fig10
\caption{\label{Fig:hvsy50}
Measured\cite{Zhu12,expt} separation versus
displacement on substrate in contact
corresponding to preceding figure.
The solid curve is extension and the dashed curve is retraction.
}
\end{figure}
%%%%%%%%%%%%%%%%%%%%%%%%%%%%%%%%%%%%%%%%%%%%%%%%%%%%%%%%%%%%%%%%%%

Figure \ref{Fig:hvsy50} shows the apparent separation versus
contact position on the substrate at this highest velocity.
In this case it is difficult to see a correlation between
the extend and the retract traces.
The spatial wave length of the noise is large,
as one might expect of temporal vibrations
at this high speed,
and they appear to dominate any topographic features.
The pronounced feature at the start of the retract trace at $y=45\,$nm
is an artefact of the turn point and the inapplicability
of the linear friction model there.
Likewise,
the pronounced feature near final contact, $y \alt 0\,$nm on retraction,
which can also be seen in Fig.~\ref{Fig:hvsy20},
is also likely an artifact
of the linear friction model, $F_y = -\mu F_z$.
In this region on retraction just prior to pull-off
there is a pronounced adhesion and $F_z < 0$.
The linear friction model is only valid, if at all, for positive loads.
In this context
it is worth mentioning that there is a modified form
of Amontons' law,
$F_y = \mu [ F_z - A ]$, where $A$ is the adhesion.\cite{Derjaguin94}
There is some experimental support for this,\cite{Stiernstedt05,Feiler00}
but it has not been explored in the present investigation.

In \S\ref{Sec:Leff} a method of extracting the drag force was given,
which was characterized by an effective length
that was expected to be somewhat less than the length of the cantilever.
The validity of the procedure could be checked from the extent
to which the effective length was independent of the drive velocity.
For velocities
$\dot z =$ 2, 20, 50,  and 80$\,\mu$m$\,$s$^{-1}$,
it was found that
$L_\mathrm{eff}=$ 83.2, 92.9, 77.7, and 88.1$\,\mu$m, respectively.
These are remarkably consistent.
There was some dependence on the choice of the base-line region.
In the worst case, $\dot z =$ 2$\,\mu$m$\,$s$^{-1}$,
reducing the range of the base-line by a factor of almost 3,
from 1.83$\,\mu$m to 0.65$\,\mu$m increased the drag length
by about 20\%.

% end comment %%%%%%%%%%%%%%%%%%%%%%%%%%%%%%%%%%%%%%%%%%%%%%%%%

%%%%%%%%%%%%%%%%%%%%%%%%%%%%%%%%%%%%%%%%%%%%%%%%%%%%%%%%%%%%%%%%%%%%%%%%%%
%
\section{Conclusion}
%
%%%%%%%%%%%%%%%%%%%%%%%%%%%%%%%%%%%%%%%%%%%%%%%%%%%%%%%%%%%%%%%%%%%%%%%%%%

This paper has investigated the causes
of non-linearity in the contact region
of atomic force microscope measurements of surface forces,
and has developed an algorithm for analyzing data
when such curved compliance occurs.
Two sources of non-linearity
---large cantilever deflection and photo-diode response---
were identified and investigated.
It was concluded that in a typical case
the non-linear photo-diode response
was the dominant contribution to the observed
non-linearity in contact.

A relatively simple algorithm for analyzing raw experimental data
was developed.
The numerical algorithm invoked
a non-linear polynomial fit to the measured voltage
in the contact region,
and was found easy to implement with a  spread sheet.

The first advantage of the algorithm is that it eliminates
the ambiguity in the choice of the contact slope
(constant compliance factor)
that occurs when one has a curved contact region.
Even when, or, more precisely, especially when,
the non-contact forces are in the linear response regime,
one still needs this calibration factor to convert the measured
photo-diode voltage to force and separation.
It significantly improves the quantitative reliability
of the atomic force microscope
to have an algorithm that gives it correctly and unambiguously.

A second advantage is that the algorithm eliminates non-physical behavior
of the analysed data in the contact region
that is an artifact of the linear analysis.
The results show that the non-linear analysis
yields reliable topographic data in the contact region
with sub-nanometer accuracy.

The complete non-linear analysis of cantilever deflection
is useful even though it turns out
that in this instance it is not required in full.
The full analysis was simplified to the linear case
but it still included the effects of cantilever tilt,
friction in contact, and torque due to the extended probe.
These are often neglected in the conventional analysis
but they are required for accurate and reliable
analysis of raw experimental data.
In particular,
a third feature of the analysis
is that it gives the relationship between
the intrinsic cantilever spring constant,
which is a material property of the cantilever
and which is the quantity usually measured by calibration procedures,
and the effective spring constant,
which is the quantity required to convert
the measured vertical cantilever deflection to a surface force.
The difference between these two can typically be 10\% or more,
and using the wrong one can lead to unacceptable errors in the measured
surface force.

A fourth feature of the data analysis algorithm
is the accounting of measurement artifacts
such as virtual deflection,
thermal drift,
cantilever drag,
and long-range surface force asymptotes.
The proper treatment and elimination of these
is automated using relatively straight forward
linear corrections.

A fifth innovation is an automated numerical procedure
for fixing the zero of separation.
The advantage of this is that it eliminates the ambiguity
that exists in the conventional analysis
and it avoids human intervention and bias.
An unambiguous and precise definition of zero
can be essential when finer details of surface forces
in the non-contact region are required.

%%%%%%%%%%%%%%%%%%%%%%%%%%%%%%%%%%%%%%%%%%%%%%%%%%%%%
\subsubsection{Reconciliation with Earlier Results}

The non-linear algorithm was applied
to measured raw atomic force microscope data
that had  previously been analyzed using
the conventional linear approach.
The validity of non-linearly analyzed data
was confirmed by the quantitative agreement
with the calculated drainage force
over a range of velocities.\cite{Zhu12}
In particular, quantitative agreement
occurred for the measured drainage adhesion,
which had not previously been analyzed in detail.

Some effort was made to test the conventional linear
analysis when the data showed a non-linear contact region
(curved compliance).
It was found that provided the correct effective spring constant was used
in the linear analysis, the measured pre-contact forces
agreed with the non-linearly analyzed data.
Obviously, if the intrinsic cantilever spring constant was used instead,
as is often the case in conventional analysis,
the non-contact force was underestimated by about 10\%.
Also, in order to get agreement,
the zero of separation had to be established using the (linear)
algorithm given here.

The slip length obtained on the basis of the present non-linear analysis
was 3$\,$nm and it was found to be independent of shear rate.
In the previous work where the data was analyzed with the
linear algorithm,\cite{Zhu12}
the slip length was found to decrease with increasing shear rate,
and to have a low shear rate limiting value of 10$\,$nm.
These qualitative and quantitative discrepancies
were attributed directly to an underestimate of the cantilever spring constant
in the previous work,\cite{Zhu12}
which is a consequence of the linear analysis used there.

In the light of the present results and analysis
one can explain the reasons for the two major discrepancies
between the present results and the earlier results:
the much larger slip length,
and the decrease in slip length with increasing shear rate
found earlier.
The earlier analysis used as the calibration factor
(the conversion factor between photodiode voltage and cantilever deflection)
the gradient of the contact extension curve measured at first contact.
This corresponds to the lowest extension contact force but the greatest
extension non-contact drainage force.
Because the contact curve is concave down due to the non-linearity
in the photodiode response (see Fig.~\ref{Fig:VvsZp}),
this value of the calibration factor \emph{underestimates}
the actual response in the non-contact region.
Hence, the deflection of the cantilever
(voltage divided by calibration factor) was \emph{overestimated}
in the earlier work.
Hence the effective spring constant used in the theory
to fit this measured deflection at large separations
was \emph{underestimated}.
(This explains the earlier $k_\mathrm{eff} = 1.5\,$N/m,\cite{Zhu12}
compared to the present $k_\mathrm{eff} = 1.68\,$N/m.)
However,
using too soft a cantilever in the theoretical calculations
of the drainage force
\emph{overestimates} the curvature (rate of increase with separation)
in the calculated deflection
compared to the actual curvature at intermediate separations.
For the given spring constant,
the only way to reduce the theoretical force so that
it agrees with the calculated force  at intermediate separations
is too increase the slip length,
which causes it to be \emph{overestimated}.
(This explains the earlier $b_0 = 10\,$nm,\cite{Zhu12}
compared to the present $b = 3\,$nm.)
However, at small separations,
where the pre-contact voltage is close to the first contact voltage,
the calibration factor that was used earlier is relatively accurate,
and the artificial compensation factors
(too small a spring constant, too large a slip length)
that worked at large and intermediate separations are not required
at small separations.
Hence for the given spring constant,
the overestimated slip length that fitted at intermediate separations
needs to be \emph{reduced} at small separations
to fit theory to the analysed measured data.
Since the shear rate increases with decreasing separation,
this explains the discrepancy between the earlier
conclusion that the slip length decreases
with increasing shear rate,\cite{Zhu12}
and the present conclusion that it is constant.
This chain of reasoning likely
applies as well to other measurements.\cite{Zhu11a,Zhu11b,Zhu12a}

Besides substantially improving
the reliability of the non-contact results,
the non-linear analysis gives
useable results in the contact region.
It was shown that reliable topographic information
could be extracted from the contact data
with sub-nanometer resolution.
The data was most reliable at low drive velocities.

A second material property
obtainable in  contact was the friction coefficient.
The value obtained here at the lowest velocity,
$\mu = 0.35$,
was consistent with values previously obtained
for Si--Si.\cite{Stiernstedt05}
However, the results were sensitive to the point in contact
at which it was measured,
or else to the range chosen if it was averaged in the contact region.
Also, they deteriorated badly as the drive velocity was increased.
It was concluded that the results
were being effected by variable drag,\cite{Zhu11a,Zhu11b}
which was neglected here.
This effect is exacerbated in high viscosity liquids
and at high drive velocities.
Fortunately, it was shown that the pre-contact forces
in the analyzed experimental data
were not sensitive to the value of the friction coefficient.
However,
to improve the performance in high viscosity liquids
of this method for measuring friction,
variable drag will have  to be accounted for.

%%%%%%%%%%%%%%%%%%%%%%%%%%%%%%%%%%%%%%%%%%%%%%%%%%%%%%%%%%%%%%%%%%%%%%%%%%
\acknowledgements
The raw data used here and originally analyzed
in Ref.~\onlinecite{Zhu12}
were measured by Liwen Zhu
under the supervision of Chiara Neto.
I thank Dr Zhu for providing the data
and for raising the issue a curved contact region.

%\newpage
%%%%%%%%%%%%%%%%%%%%%%%%%%%%%%%%%%%%%%%%%%%%%%%%%%%%%%%%%%%%%%%%%%%%%%%%%%

%%%%%%%%%%%%%%%%%%%%%%%%%%%%%%%%%%%%%%%%%%%%%%%%%%%%%%%%%%%%%%%%%%%%%%%%%%

\begin{thebibliography}{99}
\bibitem{NB1}
In the atomic force microscope modeled here,
the piezo-drive is connected to the cantilever holder
and is above the substrate.
In extension the velocity of the piezo-drive is negative,
$\dot z_\mathrm{p} < 0$,
and in contact the movement of the cantilever tip
is equal and opposite to that of the piezo-drive,
$\Delta z_\mathrm{t} = -\Delta z_\mathrm{p}$.
In other models of the atomic force microscope,
it is the substrate that is connected to the  piezo-drive,
so that $\dot z_\mathrm{p} > 0$ on extension
and in contact $\Delta z_\mathrm{t} = \Delta z_\mathrm{p}$.
These other models may be analysed with the formulae given in this paper
by negating the piezo-drive position in the raw  data,
$z_\mathrm{p} \Rightarrow -z_\mathrm{p}$.
\bibitem{expt}
The atomic force microscope was an Asylum Research Instruments, MFP-3D.
The probe was a silicon colloid, $R=10.11\,\mu$m,
the tilt angle was $\theta_0=-11^\circ$,
the substrate was a silicon wafer,
and the liquid was di-n-octylphthalate
(DOPC, viscosity $\eta \approx 52\,$mPa$\,$s).
A rectangular cantilever (NSC12/tipless (4c), Mikromasch, Estonia)
of length $L_0 = 130-20=110\,\mu$m was used.
%spring constant $k_0=1.37\,$N/m,
The measured data are taken from the same series
of experiments as those reported in Ref.~\onlinecite{Zhu12},
which may be consulted for further details.
\end{thebibliography}
\end{document}